\documentclass[twoside,11pt]{article}

\usepackage[preprint]{jmlr2e}
\usepackage{amsmath}
\usepackage{booktabs}
\usepackage{tikz}
\usepackage{mathrsfs}
\usepackage[linesnumbered,ruled,lined]{algorithm2e}
 \hypersetup{ hidelinks }

\newtheorem{assumption}{Assumption}
\DeclareMathOperator*{\argmax}{arg\,max}
\DeclareMathOperator*{\argmin}{arg\,min}
\def\PP{{\mathbb{P}}}
\def\nn{{n_1}}

\newcommand{\ind}{\rotatebox[origin=c]{90}{$\models$}}
\def\T{ {\mathrm{\scriptscriptstyle T}} }

\usepackage{lastpage}
\jmlrheading{23}{2022}{1-\pageref{LastPage}}{9/21; Revised
2/22}{8/22}{21-1120}{Baoluo Sun, Yifan Cui and Eric Tchetgen Tchetgen}

\ShortHeadings{Selective Machine Learning with an Invalid Instrumental Variable}{Sun, Cui and Tchetgen Tchetgen}
\firstpageno{1}

\begin{document}

\title{Selective Machine Learning of the Average Treatment Effect with an Invalid Instrumental Variable}

\author{\name Baoluo Sun \email stasb@nus.edu.sg \\
       \addr Department of Statistics and Data Science\\
       National University of Singapore
       \AND
       \name Yifan Cui \email cuiyf@zju.edu.cn \\
       \addr Center for Data Science\\
       Zhejiang University
       \AND
       \name Eric Tchetgen Tchetgen \email ett@wharton.upenn.edu \\
       \addr Department of Statistics and Data Science\\
       The Wharton School, University of Pennsylvania}

\editor{Victor Chernozhukov}

\maketitle

\begin{abstract}
Instrumental variable  methods have been widely used to identify causal effects in the presence of unmeasured confounding. A key identification condition known as the exclusion restriction states that the
instrument cannot have a direct effect on the outcome which is not mediated by the
exposure in view. In the health and social sciences, such an assumption is often not credible. To address this concern, we consider identification conditions of the population average treatment effect with an invalid instrumental variable which does not satisfy  the exclusion restriction, and derive the efficient influence function targeting the identifying functional under a nonparametric observed data model. We propose a novel multiply robust locally efficient estimator of the average treatment effect that is consistent in the union of  multiple parametric nuisance models, as well as a multiply debiased machine learning estimator for which the nuisance parameters are estimated using generic machine learning methods,  that effectively exploit various forms of linear or nonlinear structured sparsity in the nuisance parameter space. When one cannot be confident that any of these machine learners is consistent at sufficiently fast rates  to ensure $\surd{n}$-consistency for the average treatment effect, we introduce new criteria for selective machine learning which leverage the multiple robustness property in order to ensure small bias. The proposed methods are illustrated through extensive simulations  and a data analysis evaluating the causal effect of 401(k) participation on savings.
\end{abstract}

\begin{keywords}
Average treatment effect, Exclusion restriction, Instrumental variable, Machine learning, Multiple robustness
\end{keywords}

\section{Introduction}

One of the main concerns with drawing causal inferences from observational data is the inability to categorically rule out the existence of unobserved factors  that are associated with both the exposure and outcome variables. The instrumental variable (IV) method is  widely used in the health and social sciences for identification and estimation of causal effects under potential unmeasured confounding \citep{bowden1990instrumental, robins1994correcting, angrist1996identification,greenland2000introduction,wooldridge2010econometric,hernan2006instruments,didelez2010assumptions}.   A valid IV is a pre-exposure variable that is (a) associated with treatment, (b) independent of any unmeasured confounder of the exposure-outcome relationship, and (c) has no direct causal effect on the outcome which is not fully mediated by the exposure. While the IV approach has a longstanding tradition in econometrics going back to the original works of \cite{wright1928tariff} and \cite{goldberger1972structural} in the context of linear structural modeling, \cite{robins1994correcting}, \cite{angrist1995identification}, \cite{angrist1996identification} and \cite{heckman1997instrumental} formalized the approach under the potential outcomes framework \citep{neyman1923applications, rubin1974estimating} which allows one to nonparametrically define the causal estimands of interest and  clearly articulate assumptions needed to identify this effect; see recent reviews provided by \cite{imbens2009recent}, \cite{imbens2014},  \cite{baiocchi2014instrumental} and \cite{swanson2018partial}.  {The efficient score for the target estimands of interest in the nonparametric IV model satisfies the so called Neyman orthogonality condition \citep{neyman1,neyman2,belloni2017program, 10.1111/ectj.12097, chernozhukov2020locally}, which translates to reduced local sensitivity with respect to nuisance parameters. This allows for $\surd{n}$-consistent estimation of the causal estimands of interest even when the complexity of the nuisance parameter space  is no longer tractable by standard empirical process methods (e.g. Vapnik-Chervonenkis and Donsker classes) \citep{10.1111/ectj.12097, chernozhukov2020locally}, which represents a significant advancement in the use of machine learning methods for causal inference. }

While (b) may be ensured partly through the randomization of the IV either by design or through some natural or quasi-experiments,   the exclusion restriction (c) is not always credible  in observational studies as it requires extensive understanding of the causal mechanism by which each potential IV influences the outcome \citep{hernan2006instruments, imbens2014}. In randomized controlled studies with non-compliance, treatment assignment  may have a direct effect on the outcome if double-blinding is either absent or compromised, therefore rendering it invalid as an IV for the effects of treatment actually taken \citep{ten2008intent}.  Throughout, we shall refer to an invalid IV as a potential IV for which exclusion restriction (c) is violated. In response to this concern, there has been growing interest in the development of statistical methods
to detect and account for violation of the exclusion restriction \citep{small2007sensitivity,han2008detecting,lewbel2012using,conley2012plausibly, kolesar2015identification, bowden2016consistent, kang2016instrumental, shardell2016instrumental,wang2018sensitivity,windmeijer2018use,guo2018confidence}, primarily in a system of linear structural equation models. {To the best of our knowledge, to date there has been no published work on the population average treatment effect (ATE) as a nonparametric functional targeted with an invalid IV, which prevents the use of  data-adaptive approaches such as machine learning methods for estimation.} In this paper, we provide a novel, general set of sufficient conditions under which the ATE is nonparametrically identified despite the IV being invalid, without {\it a priori} restricting the nuisance parameters including the model for the conditional treatment effect given observed covariates. In the absence of covariates, identification and inference reduces to a setting studied recently by \citet{2019arXiv170907779T}.  Our work in this paper considerably broadens the scope of inference by allowing for potentially high dimensional covariates,  which is far more challenging than what prior literature has considered. 

For inference about the ATE, we pursue two distinct strategies for modeling the nuisance parameters: the first using standard parametric models, while the second leverages modern machine learning. In the former case we propose a multiply robust locally efficient estimator of the ATE which remains consistent under a union of multiple models, each of which restricts a separate subset of parameters indexing the observed data likelihood through low-dimensional parametric specifications. When one cannot be confident that any of these dimension-reducing models is correctly specified, we propose  flexible machine learning of nuisance parameters by selecting a learner for each nuisance parameter from an ensemble of highly adaptive candidate machine learners such as random forests, Lasso or post-Lasso and gradient boosting trees. Building upon recent work by \citet{10.1111/ectj.12097,chernozhukov2020locally} and \citet{cui2019biasaware}, the second main contribution of this paper is to introduce a novel framework for selective machine learning based on minimization of a certain cross-validated quadratic pseudo-risk which embodies the multiple robustness property. The proposed approach ensures that selection of a machine learning algorithm for a given nuisance function is made to minimize
bias of the ATE estimator associated with a suboptimal choice of machine learning algorithms to estimate the other nuisance functions. Our selective machine learning framework can be generally used for making inferences about a finite-dimensional functional defined on semiparametric models which admit multiply robust estimating functions; examples include multiply robust estimation in the context of longitudinal measurements with nonmonotone missingness \citep{vansteelandt2007estimation}, randomized trials with drop-outs \citep{tchetgen2009commentary}, statistical interactions \citep{vansteelandt2008multiply}, causal mediation analysis \citep{tchetgen2012semiparametric}, instrumental variable analysis \citep{wang2018bounded,cui2020semiparametric} and causal inference leveraging negative controls \citep{shi2020multiply}. 

The rest of the article is organized as follows.  In Section \ref{sec:framework}, we introduce the invalid IV model and provide formal identification conditions for the ATE in this setting.  We present semiparametric estimation methods in Section \ref{sec:mr}, and discuss the use of flexible machine learning of nuisance parameters in Section \ref{sec:ml}. We evaluate the finite-sample performance of these proposed methods through extensive simulation studies in Section \ref{sec:sim} and illustrate the approach with an application  to estimate the causal effect of 401(k) retirement programs on savings using data from the  Survey of Income and Program Participation in Section \ref{sec:app}. We conclude in Section \ref{sec:dis} with a brief discussion.

\section{Preliminaries}
\label{sec:framework}

\begin{figure}[!htbp]
 \includegraphics[scale=1]{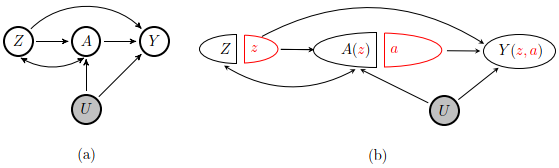}
 
	\caption{(a)  Causal Directed Acyclic Graph \citep{pearl2009causality} representing the invalid IV model within strata of measured baseline covariates,  with a bi-directed arrow between $Z$ and $A$ indicating potential unmeasured common causes of $Z$ and $A$. (b) The corresponding Single World Intervention Graph \citep{richardson2013single} with a bi-directed arrow.}
	\label{DAG:iv_model}
\end{figure}

Suppose that $(O_1,...,O_n)$ are independent and identically distributed observations of $O = (Y, A,
Z,X)$, where $Y$ is an outcome variable, $A$ is a binary treatment variable encoding the presence
$(A = 1)$ or absence of treatment $(A = 0)$, $Z$ is a binary instrument and $X$ is a set of measured baseline covariates.  { To formally define the causal estimands of interest under the potential outcomes framework \citep{neyman1923applications,rubin1974estimating}, let $Y(z,a)$ denote the potential outcome that would be observed had the instrument and exposure been set to the level $z$ and $a$ respectively, and let $A(z)$ denote the potential exposure if the instrument would take value $z$.  We  make the fundamental causal inference assumptions of (i) no interference between units and (ii) no multiple versions of the instrument and treatment; (i) and (ii) are collectively also known as the stable-unit-treatment-value assumption (SUTVA) described in \citet{rubin1980randomization}. The potential outcomes are related to the observed data via the consistency assumptions  $Y=Y(z,a)$ if $Z=z$ and $A=a$, and $A=A(z)$ if $Z=z$. We also define $Y(a)$ to be the potential outcome had only the treatment been set to level $a$, which is related via the consistency assumption $Y(a)=(1-Z)Y(0,a)+ZY(1,a):= Y(Z,a)$.

Figure \ref{DAG:iv_model}(a) gives causal graph representations \citep{pearl2009causality} of the invalid IV model considered in this paper. We assume that $U$ contains all unmeasured common causes of $A$ and $Y$, such that  conditional on $(Z,X,U)$, the effect of $A$ on $Y$ is unconfounded.
\begin{assumption}
\label{assp:1}
\begin{equation}
\begin{aligned}
\label{eq:ig}
Y(z,a)\ind A|Z=z,X,U, \text{ for all }z,a\in\{0,1\}.
\end{aligned}
\end{equation}
\end{assumption}
We will also assume that $Z$ is essentially randomized by design or through some natural experiments within strata of $X$ \citep{hernan2006instruments}.
\begin{assumption}
\label{assp:2}
\begin{equation}
\begin{aligned}
\label{eq:rand}
Y(z,a)\ind Z|U,X \text{ and } Z \ind U |X,  \text{ for all }z,a\in\{0,1\}.  
\end{aligned}
\end{equation}
\end{assumption}
Assumptions \ref{assp:1} and \ref{assp:2} may also be read (via d-separation) from the corresponding single-world intervention graph \citep{richardson2013single} in Figure \ref{DAG:iv_model}(b). The prototypical example of an invalid IV model is a randomized study where $Z$ is the treatment assignment while $A$ is the treatment actually administered, which may be influenced by some latent factors $U$  correlated with $Y$. If double-blinding is either absent or compromised, then knowledge of $Z$ may influence the post-randomization variable $Y$ directly \citep{ten2008intent}.
 
}

The ATE in the overall population $E\{Y(1)-Y(0)\}$  is arguably the causal parameter of interest in many studies  for policy questions \citep{10.2307/2291630, imbens2010better}, but
it cannot be identified under Assumptions \ref{assp:1} and \ref{assp:2} without further restrictions. Much of the invalid IV literature considered structural assumptions primarily in
multiple-IV settings $Z=(Z_1,Z_2,...,Z_p)^\T$ which imply the joint semiparametric partially linear model 
 \begin{equation}
\begin{split}
\label{eq:semiparametric1}
E(Y|A,Z,X,U)&= \boldsymbol{\theta}^\T_1 Z+\beta A +\xi_y(X,U);\\ E(A|Z,X,U)&=\boldsymbol{\theta}^\T_2 Z+\xi_a(X,U),
\end{split}
\end{equation}
indexed by the parameters $\beta \in {\rm I\!R}, \boldsymbol{\theta}_1=(\theta_{11},...,\theta_{1p})^\T\in{\rm I\!R}^{p}, \boldsymbol{\theta}_2=(\theta_{21},...,\theta_{2p})^\T\in {\rm I\!R}^p$, and  the confounding effects of the measured and unmeasured confounders on the outcome and treatment are respectively encoded by the measurable and square integrable $\xi_y(\cdot)$ and $\xi_a(\cdot)$, which remain unspecified.  Under the multivariate-IV version of Assumption \ref{assp:1} and correct specification of the partially linear model (\ref{eq:semiparametric1}), the scalar parameter $\beta$ equals the population ATE. The parameter $\boldsymbol{\theta}_1$ represents direct effects of $Z$ on $Y$. Identification of $\beta$ (or equivalently the ATE) under (\ref{eq:ig})--(\ref{eq:semiparametric1}) when $\theta_{1j}\neq 0$ for some $j\in \mathcal{J}\subseteq \{1,2,...,p\}$ has been an area of active research, generally by imposing additional restrictions on the nuisance parameter space of $\{\boldsymbol{\theta}_1,\boldsymbol{\theta}_2\}$ \citep{kolesar2015identification, bowden2016consistent, kang2016instrumental,windmeijer2018use,guo2018confidence}. 

\subsection{Nonparametric identification without exclusion restriction}
In this paper, we consider the following generalization of (\ref{eq:semiparametric1}).
\begin{assumption}
\label{assp:3}
\begin{equation}
\begin{split}
\label{eq:semiparametric2}
E(Y|A,Z,X,U)&= \theta_1(X)Z+\beta(X)A+\xi_y(X,U);\\ E(A|Z,X,U)&=\theta_2(X)Z+\xi_a(X,U),
\end{split}
\end{equation}
where $\{\beta(\cdot), \theta_1(\cdot),\theta_2(\cdot)\}$ are unknown measurable and square integrable scalar functions of the measured covariates.
\end{assumption}
Following the tradition in the IV literature \citep{robins1994correcting,angrist1995identification,angrist1996identification, heckman1997instrumental}, we focus on the canonical case of binary $A$ and $Z$; the framework can be extended readily to categorical $A$ and $Z$. The structural equation (\ref{eq:semiparametric2}) models the marginal effect of each scalar $Z$ on the outcome and the treatment which can vary with the value of observed covariates, rather than the joint effects of all available IVs, and therefore represents a significant relaxation of the restrictions in (\ref{eq:semiparametric1}). { We follow  the latent IV formulation of \cite{swanson2018partial} and formally define the no direct effect assumption or exclusion restriction as $$E\{Y(z,a)|X,U\}=E\{Y(z^{\prime},a)|X,U\} \text{ for all }z,z^{\prime},a\in\{0,1\},$$ which explicitly incorporates the unmeasured confounder $U$ \citep{dawid2003causal, didelez2010assumptions}. Under Assumptions \ref{assp:1} and \ref{assp:2}, $E\{Y(z,a)|X,U\}=E\{Y(z,a)|A=a,Z=z,X,U\}=E\{Y|A=a,Z=z,X,U\}$. Therefore $\theta_1(X)=E\{Y(1,a)|X,U\}-E\{Y(0,a)|X,U\}$ encodes the population average direct effect on the outcome within levels of $(X,U)$ for a change of the IV's value from $0$ to $1$ at each treatment level; exclusion restriction is violated in model (\ref{eq:semiparametric2}) if $\theta_1(X)\neq 0$ for at least one value in the support of $X$. In addition, under Assumption \ref{assp:1} and consistency, $E\{Y(1)-Y(0)|Z,X,U\}=E\{Y(1)|A=1,Z,X,U\}-E\{Y(0)|A=0,Z,X,U\}=E\{Y|A=1,Z,X,U\}-E\{Y|A=0,Z,X,U\}=\beta(X)$ equals the conditional ATE within levels of $(Z,X,U)$.} 

{
A design implication of (\ref{eq:semiparametric2}) is that even when $Z$ is randomized, it remains important to measure as many effect modifiers in the outcome and treatment models as possible in the hope that no residual effect modification involving $U$ remains within strata of the measured covariates $X$. We show in the Appendix that (\ref{eq:semiparametric2}) may be  relaxed so that $\{\beta(\cdot), \theta_1(\cdot),\theta_2(\cdot)\}$ varies with $U$ (albeit in restricted ways) even after controlling for $X$, a setting  also known as {\it essential heterogeneity} in the IV literature \citep{heckman2006understanding}.  Essential heterogeneity is more realistic  in a variety of settings. For example, the choice of medical treatment is likely influenced by idiosyncratic gains from alternative treatment  in the analysis of health-care decisions.  The direct effect of treatment assignment on the outcome may also be influenced by knowledge of such gains if double-blinding is either absent or compromised in randomized studies \citep{ten2008intent}. For these reasons, we focus on (\ref{eq:semiparametric2}) for identification and inference, although an alternative identification approach involving the nonlinear multiplicative model 
$$\log \{p_1(X,U)/p_0(X,U)\}=\tilde{\theta}_2(X),$$
where $p_z(X,U):=P(A=1|Z=z,X,U)$, may be used for binary treatment which rules out essential heterogeneity \citep{2019arXiv170907779T}. As pointed out by the reviewers, the function ${\theta}_2(X)$ cannot in general be variation independent of the function $\xi_a(X,U)$ if the resulting treatment conditional mean $E(A|Z,X,U)$ must remain in the unit interval. Nevertheless,  a variation independent parameterization of $E(A|Z,X,U)$ is possible such that the aforementioned dependence can be encoded in a manner compatible with our identifying assumptions, by using the odds product parameterization of \cite{ richardson2017modeling} detailed in Appendix B which is compatible with natural constraints of the data generating mechanism.
}

 Let $\beta_0(x)$,  $\mu_0(z,x):= P(A=1|Z=z,X=x)$ and $\varepsilon:=A-\mu_0(Z,X)$ denote the true conditional ATE function, treatment propensity score and the treatment regression residual respectively. In what follows the residual $\varepsilon$ serves to tease out the treatment effect via its orthogonality to direct effect component $\theta_1(X)Z$. We show in the Appendix that under Assumptions \ref{assp:2} and \ref{assp:3},
\begin{equation}
\begin{aligned}
\label{eq:ett}
E(\varepsilon Y|Z,X)=\beta_0(X)\text{Var}(A|Z,X)+\rho_0(X),
\end{aligned}
\end{equation}
 where $\text{Var}(A|Z=z,X=x):= \mu_0(z,x)\{1-\mu_0(z,x)\}$ denotes the conditional variance of $A$ within the subpopulation $\{Z=x,X=x\}$ and $\rho_0(x):= E\{\varepsilon(Y-\beta_0(X)A)|X=x\}$. The conditional covariance independence restriction $$E\{\varepsilon(Y-\beta_0(X)A)|Z,X\}=\text{Cov}\{\xi_a(X,U) ,\xi_y(X,U)|Z,X\}=\rho_0(X),$$ holds almost surely under Assumptions \ref{assp:2} and \ref{assp:3}. Therefore the function $\rho_0(\cdot)$ may be interpreted as encoding the degree of stratum-specific unmeasured confounding. Additional regularity conditions on the  observed data law are required for identification of the ATE.
 \begin{assumption}
 \label{assp:4}
The true observed data distribution lies in the interior of the nonparametric model $\mathcal{M}$ that satisfies $\text{Var}(A|Z=1,X)-\text{Var}(A|Z=0,X)\neq 0$ (heteroscedasticity), and $\pi_0(1|X):=P(Z=1|X)\in(c,1-c)$ for some $c \in (0,1/2)$ (positivity), almost surely.
  \end{assumption}
 Assumption \ref{assp:4} consists of observed data restrictions that are  empirically testable. { Heteroscedasticity has been widely used in prior works  as a source of identification in linear structural models without exclusion restrictions \citep{rigobon2003identification, klein2010estimating, lewbel2012using} and represents a strengthening of the traditional IV relevance assumption $P(A=1|Z=1,X)-P(A=1|Z=0,X)\neq 0$ in the context of binary $A$ and $Z$, as it further requires $P(A=1|Z=1,X)+P(A=1|Z=0,X)\neq 1$ to hold almost surely.} Positivity ensures that there is overlap in the distribution of baseline covariates $X$ among $Z=0$ and $Z=1$ units so that the treatment effect within each level of $X$ can be identified. Equation (\ref{eq:ett}) in conjunction with Assumptions \ref{assp:1} and \ref{assp:4} implies that 
  \begin{equation}
\begin{gathered}
\label{eq:estimand}
\text{ATE}=E\{\beta_0(X)\}=E\left\{\frac{E(\varepsilon Y|Z=1,X)-E(\varepsilon Y|Z=0,X)}{\text{Var}(A|Z=1,X)-\text{Var}(A|Z=0,X)}\right\}.
\end{gathered}
\end{equation}
 The nonparametric representation in (\ref{eq:estimand}) appears to be new in literature and has a form similar to the well-known Wald estimand as a ratio of differences between the two instrument groups. { Similar to the Wald estimand, estimation based on (\ref{eq:estimand}) may be vulnerable to bias and large variance if  $\text{Var}(A|Z,X)$ only 
weakly depends on $Z$. In recent work,  \citet{ye2021geniusmawii} proposed a measure of weak identification  relative to sample size and developed inference under a many weak invalid IVs asymptotic regime, which however requires correct specification of parametric models for all nuisance parameters.} The observed data density $P(O)$ with respect to some appropriate dominating measure factorizes as $P(Y,A|Z,X)\times P(Z|X) \times P(X)$. Evaluation of (\ref{eq:estimand}) requires knowledge of the joint density $P(Y,A|Z,X)$. As will be shown below, identification of the ATE may be established based on some but not necessarily all of these factors. We introduce the additional notation $\tau_0(z,x):= E\{Y-\beta_0(X)A|Z=z,X=x\}$ to simplify presentation for this purpose.

\setcounter{theorem}{0}

\begin{theorem}
\label{thr:identification}
Under Assumptions \ref{assp:1}--\ref{assp:4}, the ATE $\gamma:=E\{Y(1)-Y(0)\}=E\{\beta_0(X)\}$ is identified in $\mathcal{M}$ through the following three representations, each of which involves a distinct set of nuisance parameters:
\begin{description}
\item[Explicit representation (i):]  
$0=E\{\varphi_1(O;\pi_0,\mu_0)|X\}-\beta_0(X)$ almost surely, where
\begin{equation}
\begin{gathered}
\label{eq:ident1}
\varphi_1(O;\pi_0,\mu_0):=\frac{(2Z-1)\varepsilon Y }{\pi_0(Z|X)(\text{Var}(A|Z=1,X)-\text{Var}(A|Z=0,X))};
\end{gathered}
\end{equation}
\item[Implicit representation (ii):] 
$0=E\{\varphi_2(O;\pi_0,\beta_0,\tau_0)|X\}$ almost surely, where 
\begin{equation}
\begin{gathered}
\label{eq:ident2}
\varphi_2(O;\pi_0,\beta_0,\tau_0):=\frac{(2Z-1)A\{Y-\beta_0(X)A-\tau_0(Z,X)\}}{\pi_0(Z|X)};
\end{gathered}
\end{equation}
\item[Implicit representation (iii):] $0=E\{\varphi_3(O;\mu_0,\beta_0,\rho_0)|Z,X\}$ almost surely, where   
\begin{equation}
\begin{gathered}
\label{eq:ident3}
\varphi_3(O;\mu_0,\beta_0,\rho_0):=\varepsilon \{Y-\beta_0(X)A\}-\rho_0(X).
\end{gathered}
\end{equation}
\end{description}
\end{theorem}
 In particular, representation (i) provides a generalization of the results in \cite{lewbel2012using} and \cite{2019arXiv170907779T} which both rely on {\it a priori} restrictions on the functional form of the  conditional treatment effect within strata of measured covariates,
$$E\left\{\varphi_1(O;\pi_0,\mu_0)\rvert X=x\right\}=\frac{\text{Cov}\{Z,\varepsilon Y|X=x\}}{\text{Cov}\{Z,\varepsilon A|X=x\}}=\beta(x;\eta),$$
where $\eta$ is a finite-dimensional parameter.
G-estimators developed in the context  of additive and multiplicative structural mean models \citep{robins1989analysis, robins1994correcting} may be constructed based on unconditional forms of the equivalent restriction 
\begin{eqnarray}
\label{eq:g} 
\text{Cov}\{Z,\varepsilon(Y-\beta(X;\eta)A)|X=x\}=0.
\end{eqnarray}%
No such restriction is needed in representation (i). Thus in principle we can construct the plug-in estimator $\hat{\gamma}=\mathbb{P}_n\{\varphi_1(O;\hat{\pi},\hat{\mu})\}$, where  $\mathbb{P}_n$ denotes the empirical mean operator $\mathbb{P}_n \{G(O)\} = n^{-1}\sum_i G(O_i)$ and $(\hat{\pi},\hat{\mu})$ are nonparametric first-step estimators of $(\pi_0,\mu_0)$ which consists of conditional mean functions. Because we are not restricting  $\mathcal{M}$ except for regularity conditions, nonparametric estimators of $\gamma$ based on representations (i)--(iii) are in fact asymptotically equivalent with common influence function given in the following Theorem \ref{thr:efficient}.

\begin{theorem}
\label{thr:efficient}
The efficient influence function for estimating $\gamma$ in $\mathcal{M}$ is given by $$\varphi_{\textup{eff}}(O;\pi_0,\mu_0,\beta_0,\tau_0,\rho_0)-\gamma,$$ where
\begin{equation*}
\begin{aligned}
\varphi_{\textup{eff}}(O;\pi,\mu,\beta,\tau,\rho) &=\frac{(2Z-1)\left\{
\varepsilon(Y-\beta(X)A-\tau(Z,X))-\rho(X)
\right\} }{\pi(Z|X)\{\text{Var}(A|Z=1,X)-\text{Var}(A|Z=0,X)\}} +\beta\left( X\right).
\end{aligned}
\end{equation*}
Therefore, the semiparametric efficiency bound for estimating $\gamma$ in $\mathcal{M}$ is $$E\{(\varphi_{\textup{eff}}(O;\pi_0,\mu_0,\beta_0,\tau_0,\rho_0)-\gamma)^2\}.$$
\end{theorem}

In most practical settings, we anticipate that $X$ will generally be of moderate to high dimension relative
to the sample size, as analysts consider a broad collection of covariates and their functional forms in the
hope of capturing the salient features of the confounding effects. In this case, nonparametric
estimators of $\gamma$ may exhibit poor finite-sample behavior due to the curse of dimensionality \citep{robins1997toward}. Below, we describe two distinct strategies for modeling the nuisance parameters: the first
uses standard parametric models, while the second leverages modern machine learning. 

\section{Multiply robust estimation}
\label{sec:mr}

Consider the working parametric models $\{\pi(z|x;\eta_1),\mu(z,x;\eta_2),\beta(x;\eta_3),\tau(z,x;\eta_4),\rho(x;\eta_5)\}$ indexed by finite-dimensional parameters $\eta=(\eta^\T_1,\eta^\T_2,\eta^\T_3,\eta^\T_4,\eta^\T_5)^\T$. A two-step procedure to estimate the nuisance parameters is as follows:\\

{\noindent \it Procedure 1.}\\
(i) Solve the score equation $0=\mathbb{P}_n\{S(A,Z,X; \eta_1,\eta_2)\}$  to obtain $(\hat{\eta}^\T_1,\hat{\eta}^\T_2)^\T$, where $$S(A,Z,X; \eta_1,\eta_2)= \frac{\partial \log P(A,Z|X; \eta_1,\eta_2 )}{\partial ({\eta}^\T_1,{\eta}^\T_2)^\T}.$$\\
(ii)  Solve $0=\mathbb{P}_n\{ G(O;\hat{\eta}_1,\hat{\eta}_2,\eta_3,\eta_4,\eta_5 )\}$ to obtain $(\hat{\eta}^\T_3,\hat{\eta}^\T_4, \hat{\eta}^\T_5)^\T$, where
\begin{eqnarray*}
G(O;\eta):=\left[\begin{array}{c}
D_3(X)\frac{2Z-1}{\pi(Z|X;\eta_1)}\{\varepsilon(\eta_2)(Y-\beta(X;\eta_3)A-\tau(Z,X;\eta_4))-\rho(X;\eta_5)\}\\ 
D_4(X)\{Y-\beta(X;\eta_3)A-\tau(Z,X;\eta_4)\}\\
D_5(X)\{\varepsilon(\eta_2)(Y-\beta(X;\eta_3)A)-\rho(X;\eta_5)\}
\end{array}\right],
\end{eqnarray*}%
and $D_j(X)$ is a user-specified vector function of the same dimension as $\eta_j$ for $j=3,4,5$.\\

\noindent Similar to \cite{bang2005doubly, tchetgen2009doubly, sun2016semiparametric, sun2018inverse,wang2018bounded}, in the following we propose the estimator $\hat{\gamma}_{mr}= \mathbb{P}_n\{\varphi_{\text{eff}}(O;\hat{\pi},\hat{\mu},\hat{\beta},\hat{\tau},\hat{\rho})\}$ based on the form of the efficient influence function given in Theorem \ref{thr:efficient}, where $\hat{\pi}=\pi(\cdot;\hat{\eta}_1 )$, $\hat{\mu}=\mu(\cdot;\hat{\eta}_2 )$, $\hat{\beta}=\beta(\cdot;\hat{\eta}_3 )$, $\hat{\tau}=\tau(\cdot;\hat{\eta}_4 )$ and $\hat{\rho}=\rho(\cdot;\hat{\eta}_5 )$. Let $\eta^{\ast}$ denote the probability limit of $\hat{\eta}$. Because the two-step estimator $\hat{\eta}$ may be viewed as solving the joint moment equation $0=\mathbb{P}_n\{\tilde{G}(O;\eta)\}$ where $\tilde{G}(O;\gamma)=\{S^\T(A,Z,X;\eta_1,\eta_2), G^\T(O;\eta)\}^\T$ \citep{newey1994large}, the following result holds by invoking the $n^{-1/2}$ asymptotic expansion for $\hat{\eta}-\eta^{\ast}$, allowing for model misspecification \citep{white1982maximum}.

\setcounter{theorem}{0}
\begin{lemma}
\label{lemma:if}
Under standard regularity conditions for method of moments estimation  \citep{newey1994large}, $\eta^{\ast}$ is the unique solution to $E\{\tilde{G}(O;\eta)\}=0$. Furthermore, $\hat{\gamma}_{mr}$ is a consistent and asymptotically normal (CAN) estimator of ${\gamma}^{\ast}_{mr}=E\{\varphi_{\textup{eff}}(O;\eta^{\ast})\}$,
$$ \sqrt{n}(\hat{\gamma}_{mr}-{\gamma}^{\ast}_{mr})\xrightarrow[]{d}N(0,\Sigma),$$
where $\Sigma=E[\{\varphi_{mr}(O;\eta^{\ast})-{\gamma}^{\ast}_{mr}\}^2]$ and
$$\varphi_{mr}(O;\eta^{\ast})=\varphi_{\textup{eff}}(O;{\eta}^{\ast}) -E\left\{\frac{\partial  \varphi_{\textup{eff}}(O;\eta)}{\partial \eta^\T}\biggr\rvert_{\eta=\eta^{\ast}}\right\}\times \left[E\left\{\frac{\partial  \tilde{G}(O;\eta)}{\partial \eta}\biggr\rvert_{\eta=\eta^{\ast}}\right\}\right]^{-1}\tilde{G}(O;\eta^{\ast}).$$
\end{lemma}

Let ${\pi}^{\ast}=\pi(\cdot;{\eta}^{\ast}_1 )$, ${\mu}^{\ast}=\mu(\cdot;{\eta}^{\ast}_2 )$, ${\beta}^{\ast}=\beta(\cdot;{\eta}^{\ast}_3)$, ${\tau}^{\ast}=\tau(\cdot;{\eta}^{\ast}_4 )$ and ${\rho}^{\ast}=\rho(\cdot;{\eta}^{\ast}_5 )$ denote the probability limits under the (possibly misspecified) working models. Based on the three distinct sets of nuisance parameters characterized in Theorem \ref{thr:identification}, the efficient influence function has the multiple robustness property that $\gamma=E\{\varphi_{\text{eff}}(O;{\pi}^{\ast}, {\mu}^{\ast},{\beta}^{\ast},{\tau}^{\ast},{\rho}^{\ast})\}$ if at least one of the following holds: (i) $({\pi}^{\ast}, {\mu}^{\ast})=(\pi_0,\mu_0)$; (ii) $({\pi}^{\ast},{\beta}^{\ast},{\tau}^{\ast})=(\pi_0,\beta_0,\tau_0)$ and (iii) $({\mu}^{\ast},{\beta}^{\ast}, {\rho}^{\ast})=(\mu_0,\beta_0,\rho_0)$. This suggests that $\hat{\gamma}_{mr}$ is a CAN estimator of $\gamma$  under one, but not necessarily more than one, of the following three different sets of model
assumptions: 
\begin{description}

	\item[$\mathcal{M}_1$:] models for $(\pi_0,\mu_0)$ are correctly specified;
           \item[$\mathcal{M}_2$:] models for $(\pi_0,\beta_0,\tau_0)$ are correctly specified;
           \item[$\mathcal{M}_3$:] models for $(\mu_0,\beta_0,\rho_0)$ are correctly specified.
\end{description}

\begin{lemma}
\label{lemma:mul}
$\hat{\gamma}_{mr}$ is a CAN estimator of $\gamma$ in the union model $\mathcal{M}_{{{union}}}=\cup_{k=1}^3 \mathcal{M}_k$. Furthermore, $\hat{\gamma}_{mr}$ attains the semiparametric efficiency bound in $\mathcal{M}$ at the intersection submodel $\cap_{k=1}^3 \mathcal{M}_k$ where all the working models are correctly specified.
\end{lemma}
Following a theorem due to \cite{robi}, $\hat{\gamma}_{mr}$ can be shown to also attain the semiparametric efficiency bound for  $\mathcal{M}_{{{union}}}$ at the intersection submodel $\cap_{k=1}^3 \mathcal{M}_k$. Because the nuisance parameters in each of $\mathcal{M}_1$, $\mathcal{M}_2$ and $\mathcal{M}_3$ are variation independent of each other, multiply robust estimation gives the analyst three genuine opportunities to obtain valid inferences about $\gamma$, even under partial misspecification of the observed data models. 

\subsection{Comparison with existing estimators}
\label{sec:comparison}
Under the particular specification $\{\beta(x;\eta_3),\tau(z,x;\eta_4),\rho(x;\eta_5)\}=\{0,0,0\}$,  $\hat{\gamma}_{mr}$ reduces to the semiparametric plug-in estimator $\hat{\gamma}_{1}=\mathbb{P}_n\{\varphi_1(O;\hat{\pi},\hat{\mu})\}$ which is CAN only in $\mathcal{M}_{1}$. An appealing feature of $\hat{\gamma}_{1}$ is that the nuisance parameters can all be estimated in step (i) of procedure 1 without involving outcome data, and therefore mitigates potential for ``data-dredging'' exercises \citep{rubin2007design}. However, $\hat{\gamma}_{1}$ is neither multiply robust nor locally efficient.  Furthermore, $\hat{\gamma}_{mr}$ is expected to be more efficient than $\hat{\gamma}_{1}$, since the latter fails to incorporate information from $(A,Z,X)$ which may be predictive of the outcome values. Such efficiency considerations are analogous to related results on covariate adjustment in completely randomized experiments with full compliance \citep{leon2003semiparametric, davidian2005semiparametric, rubin2011targeted}.

Based on moment condition (\ref{eq:g}),  \citet{2019arXiv170907779T} proposed the covariate-adjusted ``Mendelian Randomization G-Estimation under No
Interaction with Unmeasured Selection'' (MR GENIUS) estimator $\hat{\eta}^g_{3}$ which solves
\begin{eqnarray}
\label{eq:genius} 
0= \mathbb{P}_n\left\{\frac{D_3(X)(2Z-1)\varepsilon(\hat{\eta}_2)(Y-\beta(X;\eta_3)A)}{\pi(Z|X;\hat{\eta}_1)}\right\}.
\end{eqnarray}%
 It is straightforward to verify that $\hat{\gamma}_g=\mathbb{P}_n\{\beta(X;\hat{\eta}_{3}^g)\}$ is a CAN estimator of $\gamma$ under the model assumption
 \begin{description} 
	\item[$\mathcal{M}_1^{\prime}$:] models for $(\pi_0,\mu_0, \beta_0)$  are correct.
\end{description}
{ Interestingly, under Assumptions \ref{assp:1}--\ref{assp:3} and no unmeasured confounding given $(Z,X)$, i.e., if either $U\ind A|Z,X$ or $U \ind Y|A,Z,X$, the G-estimator $\hat{\eta}_{3}$ of \citet{robins1989analysis, robins1994correcting}  solves
\begin{eqnarray}
\label{eq:g} 
0= \mathbb{P}_n\left\{D_3(X)\varepsilon(\hat{\eta}_2)(Y-\beta(X;\eta_3)A)\right\},
\end{eqnarray}%
and $\hat{\gamma}=\mathbb{P}_n\{\beta(X;\hat{\eta}_{3})\}$ is a CAN estimator of $\gamma$.
}
 Similar to standard G-estimation, a more efficient MR GENIUS estimator $\tilde{\eta}^g_{3}$ may be obtained as the joint solution to
\begin{eqnarray}
\label{eq:effgenius}  
 0= \mathbb{P}_n
\left\{\begin{array}{c}
\frac{D_3(X)(2Z-1)\varepsilon(\hat{\eta}_2)(Y-\beta(X;\eta_3)A-\tau(Z,X;\eta_4))}{\pi(Z|X;\hat{\eta}_1)}\\
D_4(X)(Y-\beta(X;\eta_3)A-\tau(Z,X;\eta_4))\\
\end{array}\right\},
\end{eqnarray}%
 where information about the association between $(Z,X)$ and $Y$ is incorporated via an additional working model for $\tau_0(\cdot)$. \citet{lewbel2012using} considered semiparametric estimation based on moment restrictions similar to (\ref{eq:effgenius}) but with nonparametric plug-ins for the nuisance parameters $(\eta_1,\eta_2)$. The resulting estimator $\tilde{\gamma}_{g}=\mathbb{P}_n\{\beta(X;\tilde{\eta}^g_{3})\}$ is doubly robust in the union model ${\mathcal{M}}^{\prime}_1 \cup {\mathcal{M}}_2$, which is in turn a submodel of $\mathcal{M}_{{{union}}}$.

\section{Flexible estimation of nuisance parameters}
\label{sec:ml}
With high-dimensional $X$, various flexible and data-adaptive statistical or machine learning  methods may be adopted to estimate the nuisance parameters $\eta_0=(\pi_0,\mu_0,\beta_0,\tau_0,\rho_0)$,  including random forests, Lasso, neural nets, boosting or their ensembles. Recent work by \cite{10.1111/ectj.12097,chernozhukov2020locally} show that $\surd{n}$-consistent estimation of  $\gamma$ is possible even when the complexity of the nuisance parameters is not tractable by standard empirical process theory (e.g. Vapnik-Chervonenkis and Donsker classes). Let $({I}_k)_{k=1}^K$ be a $K$-fold random partition of the observation indices $\{1,2,...,n\}$. For each $k$, let $\hat{\pi}(k)$, $\hat{\mu}(k)$, $\hat{\beta}(k)$, $\hat{\tau}(k)$ and $\hat{\rho}(k)$  be  learners of the nuisance parameters that are constructed using all observations not in $I_k$ based on the following two-step procedure.\\

{\noindent \it Procedure 2.}\\
(i) Obtain $\hat{\pi}(k)$ and $\hat{\mu}(k)$ by machine learning of the conditional mean functions $({\pi}_0,{\mu}_0)$. \\
(ii) Given $\hat{\pi}(k)$ and $\hat{\mu}(k)$, obtain $\hat{\beta}(k)$, $\hat{\tau}(k)$ and $\hat{\rho}(k)$  sequentially by machine learning based on the following conditional mean relationships: ${\beta}(x; \pi,\mu)=E\left\{\varphi_1(O;\pi,\mu)\rvert X=x\right\}$, $\tau(z,x; \beta)=E\{{Y}-\beta(X)A\rvert Z=z, X=x\}$ and $\rho(x;\mu,\beta)=E\{\varepsilon(\mu)({Y}-\beta(X)A)\rvert X=x\}$.\\

\noindent The cross-fitted debiased machine learning (DML) estimator of $\gamma$ is
$$\hat{\gamma}_{dml}=\frac{1}{n}\sum^K_{k=1}\sum_{i\in I_k}\varphi_{\text{eff}}(O_i;\hat{\pi}(k), \hat{\mu}(k), \hat{\beta}(k), \hat{\tau}(k) , \hat{\rho}(k)).$$
 By definition, the efficient influence function $\varphi_{\text{eff}}(O;\eta)$ satisfies the Neyman orthogonality condition \citep{neyman1,neyman2,belloni2017program, 10.1111/ectj.12097, chernozhukov2020locally}, as all first order influence functions admit second order bias \citep{robins2009quadratic}. Under general regularity conditions established by \cite{10.1111/ectj.12097, chernozhukov2020locally}, $\hat{\gamma}_{dml}$ is CAN if all the nuisance parameters are estimated with mean-squared error rates diminishing faster than $n^{-1/4}$. Such rates are achievable  for many highly data-adaptive machine learning  methods, including LASSO \citep{tibshirani1996regression}, gradient boosting trees \citep{friedman2001greedy}, random forests \citep{breiman2001random, wager2018estimation} or ensembles of these methods. { We note that in low-dimensional settings, Lemma \ref{lemma:mul} shows that $\hat{\gamma}_{mr}$ is CAN even when some of the nuisance models is misspecified by invoking the usual $n^{-1/2}$ asymptotic expansion \citep{white1982maximum}, which is not applicable when nuisance parameters are estimated via machine learning  methods. Therefore while methods such as DML and CV-TMLE  \citep{zheng2010asymptotic,van2011targeted} with machine learning remain consistent when various strict subsets of nuisance parameter learners (e.g. $\hat{\pi}$ and $\hat{\mu}$) are consistent due to the multiple robustness property of the efficient influence function $\varphi_{\text{eff}}(O;\eta)$, they generally require consistent estimation of all nuisance parameters in order to obtain valid confidence intervals.}

\subsection{Selective machine learning of multiply robust functionals}

The performance of DML estimators is intimately related to the choice of the nuisance parameter learners, even when the latter includes flexible machine learning or other nonparametric data adaptive methods. For this reason, generally one would like to learn adaptively from data and avoid choosing models {\it ex ante}. The task of model selection of parametric nuisance models was recently considered by \cite{han2013estimation}, \cite{chan2013simple}, \cite{han2014multiply}, \cite{chan2014oracle}, \cite{duan2017ensemble}, \cite{chen2017multiply} and \cite{li2020demystifying} in specific semiparametric doubly robust estimation settings.  A related strand of work is CV-TMLE which can provide notable improvements by incorporating an ensemble of semiparametric or nonparametric methods. Nonetheless, the above methods primarily focused on optimal estimation of nuisance parameters, but not bias reduction of the functional ultimately of interest. This latter task is considerably more challenging since the risk of a nonparametric functional does not typically admit an unbiased estimator and therefore may not be minimized without excessive error. \citet{cui2019biasaware} proposed a novel model selection criteria for bias reduction in estimating  nonparametric functionals of interest, based on minimization of a cross-validated empirical quadratic pseudo-risk  in the context of doubly robust estimating functions. In this paper we propose to extend their work to the multiply robust setting.

Consider the collection of candidate parametric or nonparametric learners $$\mathcal{L}=\{\hat{\eta}{(\alpha)}=\hat{\eta}{(\alpha_1,\alpha_2,\alpha_3,\alpha_4,\alpha_5)}=(\hat{\pi}_{\alpha_1}, \hat{\mu}_{ \alpha_2},\hat{\beta}_{\alpha_3},\hat{\tau}_{\alpha_4},\hat{\rho}_{\alpha_5}): 1\leq \alpha_j\leq r_j \text{ for }j=1,...,5\},$$ with probability limits $\{{\eta}{(\alpha)}={\eta}{(\alpha_1,\alpha_2,\alpha_3,\alpha_4,\alpha_5)}: 1\leq \alpha_j\leq r_j \text{ for }j=1,...,5\}$. Suppose one of the candidate learners $\hat{\eta}{(\check{\alpha})}\in \mathcal{L}$ is consistent so that ${\eta}{(\check{\alpha})}=(\pi_0,\mu_0,\beta_0,\tau_0,\rho_0)$.  The proposed procedure relies crucially on the following  two sets of mean zero implications due to the multiply robust property of the efficient influence function, that for all $1\leq \alpha_j\leq r_j$, $1\leq {\alpha}^{\prime}_j\leq r_j$, $j\in\{1,...,5\}$,

\begin{equation}
\begin{aligned}
\label{eq:mr1}
0&=E\{\bar{\varphi}_{\text{eff}}(\check{\alpha}_1,\check{\alpha}_2,\check{\alpha}_3,\check{\alpha}_4,\check{\alpha}_5)-\bar{\varphi}_{\text{eff}}(\check{\alpha}_1,\check{\alpha}_2,\alpha_3,\alpha_4,\alpha_5)\};\\
0&=E\{\bar{\varphi}_{\text{eff}}(\check{\alpha}_1,\check{\alpha}_2,\check{\alpha}_3,\check{\alpha}_4,\check{\alpha}_5)-\bar{\varphi}_{\text{eff}}(\check{\alpha}_1,\alpha_2,\check{\alpha}_3,\check{\alpha}_4,\alpha_5)\};\\
0&=E\{\bar{\varphi}_{\text{eff}}(\check{\alpha}_1,\check{\alpha}_2,\check{\alpha}_3,\check{\alpha}_4,\check{\alpha}_5)-\bar{\varphi}_{\text{eff}}(\alpha_1,\check{\alpha}_2,\check{\alpha}_3,\alpha_4,\check{\alpha}_5)\},
\end{aligned}
\end{equation}
and
\begin{equation}
\begin{aligned}
\label{eq:mr2}
0&=E\{\bar{\varphi}_{\text{eff}}(\check{\alpha}_1,\check{\alpha}_2,{\alpha}^{\prime}_3,{\alpha}^{\prime}_4,{\alpha}^{\prime}_5)-\bar{\varphi}_{\text{eff}}(\check{\alpha}_1,\check{\alpha}_2,\alpha_3,\alpha_4,\alpha_5)\};\\
0&=E\{\bar{\varphi}_{\text{eff}}(\check{\alpha}_1,{\alpha}^{\prime}_2,\check{\alpha}_3,\check{\alpha}_4,{\alpha}^{\prime}_5)-\bar{\varphi}_{\text{eff}}(\check{\alpha}_1,\alpha_2,\check{\alpha}_3,\check{\alpha}_4,\alpha_5)\};\\
0&=E\{\bar{\varphi}_{\text{eff}}({\alpha}^{\prime}_1,\check{\alpha}_2,\check{\alpha}_3,{\alpha}^{\prime}_4,\check{\alpha}_5)-\bar{\varphi}_{\text{eff}}(\alpha_1,\check{\alpha}_2,\check{\alpha}_3,\alpha_4,\check{\alpha}_5)\},
\end{aligned}
\end{equation}
where $\bar{\varphi}_{\text{eff}}(\alpha):= \varphi_{\text{eff}}(O;\eta(\alpha))$. To ease presentation, we introduce the sets $\mathscr{C}=\{1,2,3,4,5\}$, $\mathscr{C}_1=\{3,4,5\}$, $\mathscr{C}_2=\{2,5\}$ and $\mathscr{C}_3=\{1,4\}$  which index the nuisance learner components. Let $\alpha_{+k}$ denote the  counters for the nuisance learner components indexed by the elements in $\mathscr{C}_k$, e.g. $\alpha_{+1}=(\alpha_3,\alpha_4,\alpha_5)$. Similarly, let $\alpha_{-k}$ denote the counters for the nuisance learner components  indexed by the elements in $\mathscr{C}$ but not in $\mathscr{C}_k$, e.g. $\alpha_{-1}=(\alpha_1,\alpha_2)$. Then (\ref{eq:mr1}) and (\ref{eq:mr2}) may be restated more concisely as 
\begin{equation}
\begin{aligned}
\label{eq:mr3}
0=E\{\bar{\varphi}_{\text{eff}}(\check{\alpha}_{-k},\check{\alpha}_{+k})-\bar{\varphi}_{\text{eff}}(\check{\alpha}_{-k},\alpha_{+k})\},
\end{aligned}
\end{equation}
and
\begin{equation}
\begin{aligned}
\label{eq:mr4}
0=E\{\bar{\varphi}_{\text{eff}}(\check{\alpha}_{-k},{\alpha}_{+k})-\bar{\varphi}_{\text{eff}}(\check{\alpha}_{-k},\alpha^{\prime}_{+k})\},
\end{aligned}
\end{equation}
respectively, for all ${\alpha}_{+k}$, $\alpha^{\prime}_{+k}\in \mathscr{A}_{k}:=\{\alpha_{+k}:1\leq {\alpha}_j\leq r_j, j\in \mathscr{C}_k\}$, $k\in\{1,2,3\}$.{ These mean zero conditions suggest perturbing the learners indexed by $\alpha_{+k}$ and using some measure of the resulting
spread as a basis for selecting between the  learners indexed by $\alpha_{-k}$. Towards this end we introduce two different norms to define the spread or pseudo-risk. The first type is
given by the overall maximum squared bias (i.e., change in the estimated functional) induced by perturbing one distinct set of learners at
a time while holding the remaining ones fixed. For an arbitrary learner $\hat{\eta}{(\alpha^{\ast})}\in\mathcal{L}$,  we define the  minimax pseudo-risk $\mathcal{R}^{(1)}(\alpha^{\ast})=\max_{k\in \{1,2,3\}}\Lambda^{(1)}_k(\alpha),$ where 
$$ \Lambda^{(1)}_k({\alpha^{\ast}})=\max_{\substack{\alpha_{+k}\in\mathscr{A}_{k} }} [E\{\bar{\varphi}_{\text{eff}}(\alpha^{\ast}_{-k},\alpha^{\ast}_{+k})-\bar{\varphi}_{\text{eff}}(\alpha^{\ast}_{-k},\alpha_{+k})\}]^2, \text{ for }k=1,2,3.$$
The second type is given
by the sum of three maximum squared bias terms, each capturing the bias induced by
perturbing a distinct set of learners. We define the  mixed minimax pseudo-risk $\mathcal{R}^{(2)}(\alpha^{\ast})=\sum_{k=1}^3 \Lambda^{(2)}_k(\alpha^{\ast}),$ where
   $$ \Lambda^{(2)}_k({\alpha^{\ast}})=\max_{\substack{\alpha_{+k},\alpha^{\prime}_{+k} \in\mathscr{A}_{k}}} [E\{\bar{\varphi}_{\text{eff}}(\alpha^{\ast}_{-k},\alpha_{+k})-\bar{\varphi}_{\text{eff}}(\alpha^{\ast}_{-k},\alpha_{+k}^{\prime})\}]^2, \text{ for }k=1,2,3.$$
For instance, suppose we have 2 candidate learners for each of the 5 nuisance parameters, i.e., $\mathcal{L}=\{\hat{\eta}{(\alpha)}=\hat{\eta}{(\alpha_1,\alpha_2,\alpha_3,\alpha_4,\alpha_5)}: 1\leq \alpha_j\leq 2 \text{ for }j=1,...,5\}$. Then the  minimax pseudo-risk for the learner $\hat{\eta}(1,1,1,1,1)\in\mathcal{L}$  is 
 \begin{align*}
 \mathcal{R}^{(1)}(1,1,1,1,1)=&\max(\max_{\substack{\alpha_3=1,2;\alpha_4=1,2;\alpha_5=1,2 }}[E\{\bar{\varphi}_{\text{eff}}(1,1,1,1,1)-\bar{\varphi}_{\text{eff}}(1,1,\alpha_3,\alpha_4,\alpha_5)\}]^2,\\
 &\max_{\substack{\alpha_2=1,2; \alpha_5=1,2}}[E\{\bar{\varphi}_{\text{eff}}(1,1,1,1,1)-\bar{\varphi}_{\text{eff}}(1,\alpha_2,1,1,\alpha_5)\}]^2,\\ &\max_{\substack{\alpha_1=1,2; \alpha_4=1,2}}[E\{\bar{\varphi}_{\text{eff}}(1,1,1,1,1)-\bar{\varphi}_{\text{eff}}(\alpha_1,1,1,\alpha_4,1)\}]^2),
  \end{align*}
and its mixed minimax pseudo-risk is 
  \begin{align*}
  \mathcal{R}^{(2)}(1,1,1,1,1)=&\max_{\substack{\alpha_3=1,2; \alpha_4=1,2; \alpha_5=1,2\\ \alpha^{\prime}_3=1,2; \alpha^{\prime}_4=1,2; \alpha^{\prime}_5=1,2 }}[E\{\bar{\varphi}_{\text{eff}}(1,1,\alpha_3,\alpha_4,\alpha_5)-\bar{\varphi}_{\text{eff}}(1,1,\alpha^{\prime}_3,\alpha^{\prime}_4,\alpha^{\prime}_5)\}]^2 \\
 &+\max_{\substack{\alpha_2=1,2; \alpha_5=1,2 \\  \alpha^{\prime}_2=1,2; \alpha^{\prime}_5=1,2}}[E\{\bar{\varphi}_{\text{eff}}(1,\alpha_2,1,1,\alpha_5)-\bar{\varphi}_{\text{eff}}(1,\alpha^{\prime}_2,1,1,\alpha^{\prime}_5)\}]^2\\
 &+\max_{\substack{\alpha_1=1,2; \alpha_4=1,2 \\ \alpha^{\prime}_1=1,2; \alpha^{\prime}_4=1,2  }}[E\{\bar{\varphi}_{\text{eff}}(\alpha_1,1,1,\alpha_4,1)-\bar{\varphi}_{\text{eff}}(\alpha^{\prime}_1,1,1,\alpha^{\prime}_4,1)\}]^2.
 \end{align*}
 The pseudo-risks for the remaining $2^5-1$ learners in $\mathcal{L}$ are evaluated similarly. The population version of minimax learners are defined as  $\left\{\argmin_{\substack{\alpha}}\mathcal{R}^{(1)}({\alpha})\right\}$ and $\left\{\argmin_{\substack{\alpha}}\mathcal{R}^{(2)}({\alpha})\right\}$  respectively. 
}

 \subsection{Multi-fold cross-validated selection}

 {We repeatedly split the data into a training set and a validation set $S$ times to avoid overfitting in selecting the minimax learners.} For the $s$-th split where $s\in\{1,2,...,S\}$, let $\{I^m_{s}\}_{m=0,1}$ be a random bipartition of the observation indices $\{1,2,...,n\}$. We use the training sample $\{1\leq i\leq n: i\in I^s_{0}\}$ to construct the estimators $\{\hat{\eta}{(\alpha;s)}: 1\leq \alpha_j\leq r_j \text{ for }j=1,...,5\}$ based on procedure 2. For each fixed learner $\hat{\eta}{(\alpha^{\ast})}\in\mathcal{L}$, the validation sample is used to evaluate
 \begin{align*}
  \widehat{\Lambda}^{(1)}_k({\alpha^{\ast}})&=\max_{\substack{\alpha_{+k} \in\mathscr{A}_{k}}}\frac{1}{S}\sum_{s=1}^S[\mathbb{P}^1_s\{\phi_s(\alpha^{\ast};\alpha^{\ast}_{-k},\alpha_{+k})\}]^2, \quad k=1,2,3,
   \end{align*}
where  $\phi_s(\alpha;\alpha^{\prime}):= {\varphi}_{\text{eff}}(O;\hat{\eta}{(\alpha;s)})-{\varphi}_{\text{eff}}(O;\hat{\eta}{(\alpha^{\prime};s)})$ and $\mathbb{P}^{m}_s:= \frac{1}{\#\{1\leq i\leq n: i\in I^{m}_s\}}\sum_{i\in I^{m}_s}\delta_{O_i}$ for $m=0,1$, with $\delta_{O}$ denoting the Dirac measure. The empirical terms $\{ \widehat{\Lambda}^{(2)}_k({\alpha^{\ast}})\}_{k=1,2,3}$ may be evaluated similarly. We select the minimizers of the empirical pseudo-risks $\widehat{\mathcal{R}}^{(1)}(\alpha)=\max_{k\in\{1,2,3\}}\widehat{\Lambda}^{(1)}_k(\alpha)$ and $\widehat{\mathcal{R}}^{(2)}(\alpha)=\sum_{k=1}^3 \widehat{\Lambda}^{(2)}_k(\alpha)$ as our nuisance parameter learners. Let $\hat{\alpha}^{(\ell)}=\argmin_{\substack{{\alpha} }}\widehat{\mathcal{R}}^{(\ell)}({\alpha})$ for $\ell=1,2$ respectively. The two proposed selective machine learning  (SML) estimators of $\gamma$ are given by 
$$\hat{\gamma}^{(\ell)}_{sml}=\frac{1}{S}\sum_{s=1}^S\mathbb{P}^1_s\{{\varphi}_{\text{eff}}(\hat{\eta}(\hat{\alpha}^{(\ell)};s))\},$$ for $\ell=1,2$. {We provide a high-level Algorithm \ref{alg:sml} for the proposed selective machine learning  procedure in Appendix C.
 
\subsection{Excess risk bound of the proposed selectors}

 We derive risk bounds for the empirically selected minimax learners $\hat{\alpha}^{(1)}$, $\hat{\alpha}^{(2)}$ and show that their risks are not much bigger than the risks provided by the respective oracle selected learners ${\alpha}^{(1)}=\argmin_{\substack{{\alpha} }}\left\{\max_{k\in\{1,2,3\}}\dot{\Lambda}^{(1)}_k(\alpha)\right\}$ and ${\alpha}^{(2)}=\argmin_{\substack{{\alpha} }}\left\{\sum_{k=1}^3\dot{\Lambda}^{(2)}_k(\alpha)\right\}$, where
 \begin{align*} 
 \dot{\Lambda}^{(1)}_k(\alpha^{\ast})&=\max_{\substack{{\alpha}_{+k}\in\mathscr{A}_{k}}}\frac{1}{S}\sum_{s=1}^S[\mathbb{P}^1\{\phi_s(\alpha^{\ast};\alpha^{\ast}_{-k},\alpha_{+k})\}]^2;\\
 \dot{\Lambda}^{(2)}_k(\alpha^{\ast})&=\max_{\substack{ {\alpha}_{+k},{\alpha}^{\prime}_{+k}\in\mathscr{A}_{k}}}\frac{1}{S}\sum_{s=1}^S[\mathbb{P}^1\{\phi_s(\alpha^{\ast}_{-k},\alpha_{+k};\alpha^{\ast}_{-k},\alpha^{\prime}_{+k})\}]^2,
\end{align*}
and $\mathbb{P}^1$ denotes the true measure of $\mathbb{P}^1_s$. 

\setcounter{theorem}{2}
\begin{theorem}
\label{thm:risk}
Suppose the nuisance parameter learners satisfy the boundedness conditions (i) $P(c\leq \hat{\pi}_{\alpha_1}(1|X)\leq 1-c)=1$ and $P(|\widehat{\text{Var}}(A|Z=1,X;\alpha_2)-\widehat{\text{Var}}(A|Z=0,X;\alpha_2)|> 0)=1$ for $1\leq \alpha_1\leq r_1$, $1\leq \alpha_2\leq r_2$ and some $c>0$, where $\widehat{\text{Var}}(A|Z,X;\alpha_2):=\hat{\mu}_{\alpha_2}(Z,X)\{1-\hat{\mu}_{\alpha_2}(Z,X)\}$; (ii) $P(|\hat{\beta}_{\alpha_3}(X)|\leq M)=1$,  $P(|\hat{\alpha}_{\alpha_4}(Z,X)|\leq M)=1$ and $P(|\hat{\rho}_{\alpha_5}(X)|\leq M)=1$ for $1\leq \alpha_3\leq r_3$, $1\leq \alpha_4\leq r_4$, $1\leq \alpha_5\leq r_5$ and some $M>0$. Then we have that  
\begin{align*}
&\mathbb{P}^0\left\{\widetilde{\mathcal{R}}^{(1)}({\hat{\alpha}^{(1)}})\right\}\leq 
(1+2\epsilon)\mathbb{P}^0\left\{\bar{\mathcal{R}}^{(1)}({{\alpha}^{(1)}})\right\} \\
&+\left(\frac{1+\epsilon}{n^{1/q}}\right)\left(\frac{1+\epsilon}{\epsilon}\right)^{(2-q)/q}C\log\left\{1+(r_3r_4r_5)^2(r_2r_5)^2(r_1r_4)^2\right\},
\end{align*}
for any $\epsilon>0$, $1\leq q\leq 2$, and some constant $C$, where $\mathbb{P}^0$ denotes the expectation with respect to training data, \begin{align*}
\widetilde{\mathcal{R}}^{(1)}({\hat{\alpha}^{(1)}})=  \max_{\substack{k\in\{1,2,3\}}} \frac{1}{S}\sum_{s=1}^S [\mathbb{P}^1\{\phi_s(\hat{\alpha}^{(1)}; \hat{\alpha}^{(1)}_{{-k}},\tilde{\alpha}_{+k})\}]^2;\\  \bar{\mathcal{R}}^{(1)}({{\alpha}^{(1)}})= \max_{\substack{k\in\{1,2,3\}}} \frac{1}{S}\sum_{s=1}^S [\mathbb{P}^1\{\phi_s({\alpha}^{(1)}; {\alpha}^{(1)}_{{-k}},\bar{\alpha}_{+k})\}]^2,
\end{align*}
and for $k=1,2,3$,
\begin{align*}
\tilde{\alpha}_{+k}&=\argmax_{{\alpha}_{+k}\in \mathscr{A}_k} \frac{1}{S}\sum_{s=1}^S[\mathbb{P}^1_s\{\phi_s(\hat{\alpha}^{(1)}; \hat{\alpha}^{(1)}_{-k},{\alpha}_{+k})\}]^2;
  \bar{\alpha}_{+k}=\argmax_{{\alpha}_{+k}\in \mathscr{A}_k}\frac{1}{S}\sum_{s=1}^S[\mathbb{P}^1\{\phi_s({\alpha}^{(1)}; {\alpha}^{(1)}_{{-k}},{\alpha}_{+k})\}]^2.
\end{align*}
 Analogous results hold for the mixed minimax selected learner,
 \begin{align*}
&\mathbb{P}^0\left\{\widetilde{\mathcal{R}}^{(2)}({\hat{\alpha}^{(2)}})\right\}\leq 
(1+2\epsilon)\mathbb{P}^0\left\{\bar{\mathcal{R}}^{(2)}({{\alpha}^{(2)}})\right\} \\
&+\left(\frac{1+\epsilon}{n^{1/q}}\right)\left(\frac{1+\epsilon}{\epsilon}\right)^{(2-q)/q}\sum_{k=1}^3 C_k\log\left[1+(r_3r_4r_5)^{\{1+I(k=1)\}}(r_2r_5)^{\{1+I(k=2)\}}(r_1r_4)^{\{1+I(k=3)\}}\right] ,
\end{align*}
for any $\epsilon>0$, $1\leq q\leq 2$, and some constants $C_1$, $C_2$ and $C_3$, where $I(\cdot)$ is the indicator function,
\begin{align*}
\widetilde{\mathcal{R}}^{(2)}({\hat{\alpha}^{(2)}})= \frac{1}{S}\sum_{k=1}^3  \sum_{s=1}^S [\mathbb{P}^1\{\phi_s(\hat{\alpha}^{(2)}_{{-k}},\tilde{\alpha}_{+k}; \hat{\alpha}^{(2)}_{{-k}},\tilde{\alpha}^{\prime}_{+k})\}]^2;\\ 
\bar{\mathcal{R}}^{(2)}({{\alpha}^{(2)}})= \frac{1}{S}\sum_{k=1}^3  \sum_{s=1}^S [\mathbb{P}^1\{\phi_s({\alpha}^{(2)}_{{-k}},\bar{\alpha}_{+k}; {\alpha}^{(2)}_{{-k}},\bar{\alpha}^{\prime}_{+k})\}]^2,
\end{align*}
and for $k=1,2,3$,
\begin{align*}
(\tilde{\alpha}_{+k},\tilde{\alpha}^{\prime}_{+k})&=\argmax_{{\alpha}_{+k},{\alpha}^{\prime}_{+k}\in \mathscr{A}_k} \frac{1}{S}\sum_{s=1}^S[\mathbb{P}^1_s\{\phi_s(\hat{\alpha}^{(2)}_{-k},{\alpha}_{+k} ; \hat{\alpha}^{(2)}_{-k},{\alpha}^{\prime}_{+k})\}]^2;\\
(\bar{\alpha}_{+k},\bar{\alpha}^{\prime}_{+k})&=\argmax_{{\alpha}_{+k},{\alpha}^{\prime}_{+k}\in \mathscr{A}_k} \frac{1}{S}\sum_{s=1}^S[\mathbb{P}^1\{\phi_s({\alpha}^{(2)}_{-k},{\alpha}_{+k} ; {\alpha}^{(2)}_{-k},{\alpha}^{\prime}_{+k})\}]^2.
\end{align*}
\end{theorem}
\noindent The bound given in Theorem \ref{thm:risk} extends the risk bound established  in  \citet{cui2019biasaware} to the multiply robust setting, and shows that  the error incurred by the empirical risk is of order $n^{-1}$ for any fixed $\epsilon$ if $q=1$. Therefore, although the proposed selection procedure is able to incorporate both parametric and nonparametric candidate learners, it is  of most interest in machine learning  settings  where the pseudo-risk can be of order substantially larger than $O(n^{-1})$, so that the error made in selecting the cross-validated minimax learner is negligible relative to its risk and the proposed selector performs nearly as well as a oracle
selector with access to the true pseudo-risk. It is also of interest in such settings to compare  the proposed SML estimators with machine learning  estimators using ensemble methods such as super learner \citep{van2007super} which selects through  cross-validation the optimal combination from a library of candidate learners to estimate each nuisance parameter separately; we investigate their empirical performances via a simulation study in the next section.

The proposed approach is completely agnostic as to whether the collection $\mathcal{L}$ includes a consistent  learner of all the nuisance parameters. Indeed if none of them are consistent there is no estimator of $\gamma$ that can still be consistent, and the proposed  approach is mostly geared towards identifying the learner that minimizes the minimax pseudo-risks for a given data set. Standard machine learning methods such as DML do not have this built-in data-adaptive feature. To illustrate the implications of this selection procedure, we note that the bias of a DML estimator of $\gamma$ evaluated with the nuisance parameter learner $\hat{\eta}(\alpha)=(\hat{\pi}_{\alpha_1}, \hat{\mu}_{ \alpha_2},\hat{\beta}_{\alpha_3},\hat{\tau}_{\alpha_4},\hat{\rho}_{\alpha_5})$ chosen {\it ex ante} is typically of the order  $$O_p\left\{{n}^{-1/2}+||\hat{\pi}_{\alpha_1}-\pi_0||_2||\hat{\rho}_{\alpha_5}-\rho_0||_2+||\hat{\mu}_{ \alpha_2}-\mu_0||_2\left(||\hat{\beta}_{\alpha_3}-\beta_0||_2+||\hat{\tau}_{\alpha_4}-\tau_0||_2\right) \right\},$$
 which  depends crucially  on products of the learners' estimation errors. Because  $\dot{\Lambda}^{(2)}_1(\alpha)$ captures the maximum squared bias in the estimated functional induced by
perturbing only the learners indexed by $(\alpha_3,\alpha_4,\alpha_5)$, its minimizer corresponds to learners indexed by $(\alpha_1,\alpha_2)$ with smallest bias. \citet{cui2019biasaware} provided formal proof of a related result. The  mixed minimax pseudo-risk  represents a natural extension of this idea as the sum of three maximum squared bias terms, each capturing the bias induced by perturbing a distinct set of learners. Due to the dependence across cross-validation samples, formal machine learning  post selection inference is challenging and the subject of ongoing research.
}

\section{Simulation studies}
\label{sec:sim}
In this section, we investigate the finite-sample properties of the proposed estimators  under a variety of settings. Baseline covariates $X=(X_1,...,X_5)^T$ are generated from independent standard uniform distributions. We consider the functional form $X^{\star}_k=[1+\exp\{-20(X_k-.5)\}]^{-1}$ for $k=1,...,5$. The unmeasured confounder $U$ is generated from a truncated normal distribution in the interval $[-.5,.5]$ with mean 0 and variance $.25+.5X^{\star}_1+.15X^{\star}_2-.1X^{\star}_3-.1X^{\star}_4+.1X^{\star}_5$. Conditional on $(U,X)$, the invalid instrument $Z$, treatment $A$ and outcome $Y$ are generated from the models
\begin{align*}
P(Z=1|U,X)&=\{1+\exp(-0.8-X^{\star}_1+.2X^{\star}_2+.2X^{\star}_3+.2X^{\star}_4-.1X^{\star}_5)\}^{-1};\\
P(A=1|Z,U,X)&=\{1+\exp(2-1.5Z-.6X^{\star}_1+.2X^{\star}_2+.2X^{\star}_3+.1X^{\star}_4-.1X^{\star}_5)\}^{-1}+\kappa_1 U;\\
E(Y|A,Z,U,X)&=-2+(2X^{\star}_1+.5X^{\star}_2+.5X^{\star}_3) A+2X^{\star}_1+.5X^{\star}_2+.2X^{\star}_3+.1X^{\star}_4+.1X^{\star}_5\\
&\phantom{=}-2Z+\kappa_2 U,
\end{align*}
where $(\kappa_1,\kappa_2)=(.1,1)$ and the outcome error term followed standard normal distribution. We are interested in estimating $\gamma=E(2X^{\star}_1+.5X^{\star}_2+.5X^{\star}_3)=1.5$ based on the generated data for $(Y,A,Z,X)$.

\subsection{Semiparametric estimators}

We implement the five semiparametric estimators $\hat{\gamma}$, $\hat{\gamma}_{1}$, $\hat{\gamma}_{g}$, $\tilde{\gamma}_{g}$ and $\hat{\gamma}_{mr}$  using  the R package \texttt{nleqslv} \citep{hasselman2018package}, and evaluate their performances in situations where some models may be misspecified. A particular working model is misspecified when the quadratic functional form  $X^{\star\star}_k=(X_k-0.5)^2$ is used in place of $X^{\star}_k$, $k=1,...,5$. Specifically, we report results from the following four scenarios:
\begin{itemize} 
	\item[$\mathcal{S}_0$:] All models are correctly specified;
	\item[$\mathcal{S}_1$:] models for $(\pi_0,\mu_0)$  are correct, but models for $(\delta_0,\tau_0,\rho_0)$ are misspecified;
	\item[$\mathcal{S}_2$:] models for $(\pi_0,\beta_0,\tau_0)$ are correct, but models for $(\mu_0,\rho_0)$ are misspecified;
	\item[$\mathcal{S}_3$:]   models for $(\mu_0,\beta_0,\rho_0)$  are correct, but models for $(\pi_0,\tau_0)$ are misspecified.
\end{itemize}
Table \ref{table1} summarizes the results  based on 1000 repeated simulations with sample size $n=2000$ or $4000$. Standard errors are obtained using the empirical sandwich estimator for generalized method of moments \citep{newey1994large}.  {The g-estimator $\hat{\gamma}$ which does not account for unmeasured confounding shows notable bias relative to its standard error, with coverage below nominal level in all scenarios.} In agreement with theory, $\hat{\gamma}_{1}$ has negligible bias and coverage proportions close to nominal levels in scenarios $\{\mathcal{S}_j\}_{j=0,1}$, $\hat{\gamma}_{g}$ only in $\mathcal{S}_0$, $\tilde{\gamma}_{1}$  in  $\{\mathcal{S}_j\}_{j=0,2}$, and $\hat{\gamma}_{mr}$ in  $\{\mathcal{S}_j\}_{j=0,1,2,3}$, confirming its multiple robustness property.  The estimators  $\hat{\gamma}_{mr}$ and $\hat{\gamma}_{1}$ perform similarly to each other in terms of absolute bias, variance and coverage in $\mathcal{S}_1$, but $\hat{\gamma}_{mr}$ yields smaller variance than $\hat{\gamma}_{1}$ in $\mathcal{S}_0$ where all models are correct.

\begin{table}
\caption{Summary of results for semiparametric estimation of $\gamma$. The result of each scenario includes two rows, of which the first stands for $n=2000$, and the second for $n=4000$.} \label{table1}  \vspace{-.1in}

\begin{center}
\begin{tabular}{crrrrr}			\toprule

	 & \multicolumn{5}{c}{Estimator}\\		
			\cmidrule(lr){2-6}
	 &  $\hat{\gamma}$   &  $\hat{\gamma}_{1}$& $\hat{\gamma}_{g}$ &$\tilde{\gamma}_{g}$ & $\hat{\gamma}_{mr}$\\
			\midrule
			&  \multicolumn{5}{c}{$\mathcal{S}_0$}\\
 Bias &.033 &$-.024$ & $.100$ & $-.014$& $-.016$ \\
        & .031&$ .014$ & $.051$ & $-.003$& $-.002$ \\
$\sqrt{\text{Var}}$ & .056& .380 & .408 & .196 & .205 \\
                               &.041 & .269 & .199 & .141 & .145 \\
Cov95 &.911& .966 & .994 & .975 & .978 \\
          &.867 & .941 & .961 & .958 & .960 \\[.1in]
			&  \multicolumn{5}{c}{$\mathcal{S}_1$}\\
 Bias &.146& $-.024$ & $-.057$ & $-.255$& $-.098$ \\
       &.135 & $.014$ & $-.144$ & $-.214$& $-.017$ \\
$\sqrt{\text{Var}}$ &.060& .380 &   .611 & .359 & .421 \\
                            &  .042  & .269 &   .310 & .241 & .273 \\
Cov95 &.325& .966 & .981 & .931 & .972 \\
           &.101& .941 & .929 & .865 & .952 \\[.1in]
			&  \multicolumn{5}{c}{$\mathcal{S}_2$}\\
 Bias& .424 & $.267$ & $.275$ & $-.012$& $-.013$ \\
       &.417 & $.304$ & $.698$ & $-.002$& $-.002$ \\
$\sqrt{\text{Var}}$& .115& .336 & .765 & .186 & .191 \\
                               &.081 & .240 & .360 & .133 & .136 \\
Cov95 &.017& .845 & .883 & .978 & .977 \\
          &.000 & .736 & .481 & .957 & .959 \\[.1in]
			&  \multicolumn{5}{c}{$\mathcal{S}_3$}\\
 Bias & .033&  .664  & .242   &   .024     & $-.077$ \\
        & .031& $.679$ & $.112$ & $.019$& $-.014$ \\
$\sqrt{\text{Var}}$& .056& .344        &     1.504    &   .911      &  .343 \\
                              & .041 & .249 & .308 & .214 & .213 \\
Cov95 &.911&   .508      &  .988       & .985         & .991          \\
         & .867 & .162 & .981 & .990 & .962 \\
			\bottomrule 
\end{tabular}\\[.1in]
\parbox{1\textwidth}{\small Note: 
Bias and $\sqrt{\text{Var}}$ are the Monte Carlo bias and standard deviation of the points estimates, and   Cov95 is the coverage proportion of
the 95\% confidence intervals, based on 1000 repeated simulations. Outlier in one run has been removed in computation of results for $\hat{\gamma}_{g}$.}
\end{center}  \vspace{-.1in}
\end{table}

\subsection{Machine learning estimators}

{We implement the DML estimators $\hat{\gamma}_{\text{LASSO}}$, $\hat{\gamma}_{\text{RF}}$, $\hat{\gamma}_{\text{GBM}}$ and $\hat{\gamma}_{\text{SL}}$ with covariates $X$ and $K=2$, whereby the nuisance parameters were estimated  with (i) LASSO \citep{tibshirani1996regression, friedman2010regularization}, (ii) classification or regression random forests \citep{breiman2001random, liaw2002classification, malley2012probability}, (iii) gradient boosting machines \citep{friedman2001greedy} or the ensemble method super learner  based on a library
consisting of (i), (ii) and (iii), using the R packages \texttt{glmnet} \citep{glmnet}, \texttt{ranger} \citep{ranger}, \texttt{gbm} \citep{gbm} or \texttt{SuperLearner} \citep{polley2021package} respectively.} In addition, we implement the proposed SML estimators with covariates $X$ by minimizing the empirical quadratic pseudo-risks over the candidate learners $\{\hat{\eta}(\alpha): 1\leq \alpha_j\leq 3 \text{ for }j=1,...,5\}$ with covariates $X$ and $S=2$, whereby each learner component is based on (i), (ii) or (iii). Table \ref{table2} summarizes the results  based on 1000 repeated simulations  with sample size $n=2000$ or $4000$.   Because the functional form of the regressors is misspecified, $\hat{\gamma}_{\text{LASSO}}$ has noticeable bias, although it has the smallest Monte Carlo standard error. The bias of $\hat{\gamma}_{\text{RF}}$ decreases with increasing sample size, and becomes negligible at $n=4000$. The DML estimator $\hat{\gamma}_{\text{GBM}}$ has considerably large bias
due to outliers when $n =2000$, but its bias decreases when $n=4000$. Remarkably, without access to the true underlying data generating mechanism or the performance of individual DML estimators, the proposed SML estimators  nearly attain the minimum absolute bias at $n=4000$.  The mixed minimax SML estimator $\hat{\gamma}^{(2)}_{sml}$ tends to be more efficient than $\hat{\gamma}^{(1)}_{sml}$, in agreement with previous simulation results for doubly robust functionals \citep{cui2019biasaware}.
 
 \begin{table}
  {
\caption{Summary of results for machine learning estimation of $\gamma$. The result of each scenario includes two rows, of which the first stands for $n=2000$, and the second for $n=4000$.} \label{table2}  \vspace{-.1in}

\begin{center}
\begin{tabular}{cccrrrrr}			\toprule

	 & \multicolumn{4}{c}{Estimator}\\		
			\cmidrule(lr){2-7}
	    &  $\hat{\gamma}_{\text{LASSO}}$ & $\hat{\gamma}_{\text{RF}}$&$\hat{\gamma}_{\text{GBM}}$ &$\hat{\gamma}_{\text{SL}}$& $\hat{\gamma}^{(1)}_{sml}$ & $\hat{\gamma}^{(2)}_{sml}$\\
			\midrule
\phantom{-}
 Bias & $0.111$ & $-0.040$ & 129.583 & $-1.704$& $-0.760$ & $0.089$ \\
        & $ 0.088$ & $-0.020$ & $-0.841$ & 0.063 & $\phantom{-}0.010$ & $0.049$  \\
$\sqrt{\text{MSE}}$ &1.658 & $\phantom{-}6.118$ & $3750.165$&45.441 & 25.660 & 1.442  \\
                                & 0.226 & $\phantom{-}0.750$ & 39.048 & 1.344& \phantom{-}1.495&  0.473\\
			\bottomrule 
\end{tabular}\\[.1in]
\parbox{1\textwidth}{\small Note: Bias and $\sqrt{\text{MSE}}$ are the Monte Carlo bias and root mean square error of the points estimates based on 1000 repeated simulations. }
\end{center}  \vspace{-.1in}
}
\end{table}

\section{Application}
\label{sec:app}
The causal relationship between 401(k) retirement programs and savings has been a subject of considerable interest in economics \citep{poterba1995401,poterba1996retirement, abadie2003semiparametric, benjamin2003does,doi:10.1162/0034653041811734}. The main concern with  causal inference based on observational data is that program participation is not randomly assigned, but rather are self-selected by individuals. Potential unmeasured confounders $U$ such as individual preferences may affect both program participation and savings. Thus, estimation of the effects of tax-deferred retirement programs may be biased even after controlling for observed covariates \citep{abadie2003semiparametric}. \cite{poterba1995401} proposed 401(k) eligibility as an instrument for program participation. If individuals made employment decisions based on income and within jobs classified by income categories, whether or not a firm offers a 401(k) plan can essentially be viewed as randomized conditional on income and other measured covariates since eligibility is determined by employers. However, 401(k) eligibility may also interact  with some unobserved heterogeneity at the firm's level in  influencing savings other than through 401(k) participation \citep{engen1996illusory}.

  In this section, we illustrate the proposed methods by reanalyzing the data from the 1991 Survey of Income and Program Participation ($n=9,915$) used in \cite{doi:10.1162/0034653041811734}. The treatment variable $A$ is a binary indicator of participation in a 401(k) plan and $Z$ is a binary indicator of 401(k) eligibility. In this dataset, 37\% are eligible for 401(k) programs and 26\% participated. The outcomes of interest are net financial assets and net non-401(k) financial assets  in 1991 (US dollars). Following \cite{poterba1995401}, \cite{ benjamin2003does} and \cite{doi:10.1162/0034653041811734}, the vector of measured covariates $X$ includes an intercept, family size, indicators for marital status, two-earner status, defined benefit pension status, IRA participation status, homeownership status, four categories of number of years of education, five categories of age and seven income categories. We  consider as benchmarks the two models
\begin{equation}
\begin{aligned}
\label{eq:outstruct2}
E(Y|A,Z,X,U)&=\beta(X) A+\xi_y(X);\\
E(Y|A,Z,X,U)&=\beta(X)A+\xi_y(X,U), 
\end{aligned}
\end{equation}
which are special cases of the outcome structural equation in Assumption \ref{assp:3}. The former model holds when the effect of $A$ on $Y$ is unconfounded conditional on $X$, and yields the observed data model $E(Y|A,X)=\beta(X) A+\tau(X)$. If we specify the parametric models $\beta(X;\eta_3)=\eta_3$ and $\tau(X;\eta_4)=\eta^\T_4 X$, the parameters $(\eta_3,\eta_4)$  indexing the conditional mean model $E(Y|A,X;\eta_3,\eta_4)$ may be estimated via  ordinary least squares. On the other hand, the latter model in (\ref{eq:outstruct2}) holds in the presence of unmeasured confounding if $Z$ is a valid instrument that satisfies exclusion restriction. The observed data model $E(Y|Z,X)=\beta(X)\mu(Z,X)+\tau(X)$ may be estimated using two-stage instrumental variable estimation under an additional model for the propensity score, which we specify as $\mu(Z,X;\eta_2)=\{1+\exp(-\eta^\T_2 (Z,X^\T)^\T)\}^{-1}$. We denote the resulting semiparametric ATE estimators as $\hat{\gamma}_{ols}$ and $\hat{\gamma}_{tsiv}$ respectively. For comparison, we implement the proposed semiparametric estimators under the same parametric models for $\delta(X)$ and $\mu(Z,X)$, as well as the additional models $\pi(1|X; \eta_1)=\{1+\exp(-\eta^\T_1 X)\}^{-1}$, $\tau(Z,X;\eta_4)=\eta^\T_4 (Z,X^{\T})^\T $ and $\rho(X;\eta_5)=\eta^\T_5 X$. The results are summarized in Table \ref{tab:result}.
 
 \begin{table} 
   {
		\begin{center}
			\caption{Estimates of the causal effect of 401(k) program participation on  savings in 1991 (US dollars).} 
			\bigskip
			\begin{tabular}{lcccccccccccc}
				\toprule
				  &Average treatment effect&Direct effect of 401(k) eligibility\\
				\midrule
				&\multicolumn{2}{c}{Net financial assets}\\[2pt] 
				$\hat{\gamma}_{ols}$ & $14517\pm 2743\phantom{^{\natural}}$ &  \\ 
				$\hat{\gamma}_{tsiv}$ & $13491\pm 4490\phantom{^{\natural}}$ &  \\ 
				$\hat{\gamma}_{1}$ & $13248\pm 5524\phantom{^{\natural}}$&\\
				$\hat{\gamma}_{g}$  &$13669\pm4216\phantom{^{\natural}}$&  \\ 
				$\tilde{\gamma}_{g}$ & $13083\pm4199\phantom{^{\natural}}$  & $2.04\pm4189$ \\
				$\hat{\gamma}_{mr}$ & $13610\pm2648\phantom{^{\natural}}$ &  $3.35\pm3879$ \\
				$\hat{\gamma}_{SL}$ & $13637\pm 5837^{\natural}$ &\\
				$\hat{\gamma}^{(1)}_{sml}$ & $13952\pm 7181^{\natural}$&\\
				$\hat{\gamma}^{(2)}_{sml}$ & $14136\pm6032^{\natural}$&\\[2pt] 
				&\multicolumn{2}{c}{Net non-401(k) financial assets}\\ [2pt]
				$\hat{\gamma}_{ols}$ & $\phantom{-}673\pm2571\phantom{^{\natural}}$ & \\ 
				$\hat{\gamma}_{tsiv}$ & $-546\pm4226\phantom{^{\natural}}$ & \\ 
				$\hat{\gamma}_{1}$ & $1129\pm5351\phantom{^{\natural}}$& \\
				$\hat{\gamma}_{g}$  &$1618\pm4077\phantom{^{\natural}}$& \\ 
				$\tilde{\gamma}_{g}$ & $1838\pm4058\phantom{^{\natural}}$ & $-1436\pm4092$ \\
				$\hat{\gamma}_{mr}$ & $1294\pm2411\phantom{^{\natural}}$ & $-1436\pm3766$\\
				$\hat{\gamma}_{SL}$ & $1368\pm 5637^{\natural}$ &\\
				$\hat{\gamma}^{(1)}_{sml}$ & $1933\pm 6391^{\natural}$&\\
				$\hat{\gamma}^{(2)}_{sml}$ & $1853\pm 6063^{\natural}$&\\						
				\bottomrule
			\end{tabular}\\[.1in]
			\parbox{1\textwidth}{\small Note: Point estimate $\pm$ 2$\times$standard error.  Following \cite{10.1111/ectj.12097}, the median estimate  out of 100 repetitions are reported for the machine learning estimators to mitigate the finite-sample impact of any particular sample splitting realization. {$\natural$ Nominal standard errors obtained based on the empirical efficient influence functions evaluated under selected learners.}}
			\label{tab:result}
		\end{center}
		}
	\end{table}
	
\subsection{Effect of 401(k) participation on  net financial assets}
The point estimate of $\hat{\gamma}_{tsiv}$ is noticeably smaller than that of $\hat{\gamma}_{ols}$, which  is consistent with the results in  \cite{doi:10.1162/0034653041811734} and suggests that unmeasured confounding generates an upward-biased estimate of the effect of 401(k) participation on savings. The proposed estimators yield significant and uniformly positive point estimates that are close to that of $\hat{\gamma}_{tsiv}$, which provide further evidence that 401(k) participation increases net financial assets even when exclusion restriction may be implausible. This similarity between $\hat{\gamma}_{tsiv}$ and the proposed estimators is due to the near zero point estimate for the coefficient corresponding to the main effect of $Z$ in $\eta_4$, which encodes the direct effect of  401(k) eligibility on net financial assets. {Compared to the semiparametric estimators, the proposed SML estimators  implemented with covariates $X$ and whereby each learner component is based on LASSO, random forests or gradient boosting machines yield similar positive effect estimates, but with noticeably larger nominal standard errors. This difference in efficiency between semiparametric and nonparametric data-adaptive estimation agrees with the simulation results in Section \ref{sec:sim}.}

\subsection{Effect of 401(k) participation on  net non-401(k)  financial assets} 
 Consistent with the findings in  \cite{doi:10.1162/0034653041811734}, the  point estimate of $\hat{\gamma}_{tsiv}$ is negative and noticeably lower than that of  $\hat{\gamma}_{ols}$. On the other hand, the proposed estimators yield uniformly positive point estimates, which suggests that 401(k) participation does not crowd out  non-401(k) savings, although the estimates are not statistically significant. This difference may be partially explained by the noticeably negative point estimate for the direct effect of 401(k) eligibility on  non-401(k) savings, indicating asset substitution in 401(k) eligible firms which obscured the effect of 401(k) participation.   This may happen, for example, if 401(k) eligible employees have access to improved financial education
and advice on 401(k) plans at the firm level which encourage  asset substitution.

\section{Discussion}
 \label{sec:dis}

 There are several improvements and extensions for future work. The finite sample performance of the proposed semiparametric estimators can be improved in terms of efficiency \citep{tan2006distributional, tan2010nonparametric} and bias \citep{vermeulen2015bias}. Efficiency can also potentially be improved by incorporating {\it a priori} knowledge such as degree of exclusion restriction violation or using IVs that are known to be valid in conjunction with the invalid ones.  Lastly, multiple invalid weak IVs can be incorporated by adopting the generalized method of moments approach \citep{Newey:2009aa, ye2021geniusmawii}.


\acks{The authors would like to thank the Action Editor and two anonymous referees for many
constructive comments which greatly improved the paper. Baoluo Sun's work is supported by the National University of Singapore Start-Up Grant R-155-000-203-133. Eric Tchetgen Tchetgen's work is funded by NIH grants R01AI27271, R01CA222147, R01AG065276 and R01GM139926. }


\newpage

\appendix
\section*{Appendix A. Proofs}
\label{app:theorem}
\renewcommand{\theequation}{A\arabic{equation}}
\setcounter{equation}{0} 


\subsection*{Proof of Theorem \ref{thr:identification}}

 \cite{2019arXiv170907779T} considered the following generalization of Assumption \ref{assp:3}  in which both the treatment effects and treatment choice may depend on $U$:
 \\
 
 {\noindent \bf Assumption 3$^\prime$}
\begin{equation*}
\begin{split}
E(Y|A,Z,X,U)&= \theta_1(X,U)Z+\beta(X,U)A+\xi_y(X,U);\\ E(A|Z,X,U)&=\theta_2(X,U)Z+\xi_a(X,U),
\end{split}
\end{equation*}
{\it where $\{\beta(\cdot), \theta_1(\cdot),\theta_2(\cdot)\}$ are unknown measurable and square integrable functions of both measured and unmeasured confounders.}
\\

\noindent This situation is also known as {\it essential heterogeneity} in the econometrics literature \citep{heckman2006understanding}. Following the proof of Lemma 3.1 in \cite{2019arXiv170907779T}, the covariate-specific equality 
\begin{equation}
\begin{aligned}
\label{eq:decomposition}
E(\varepsilon Y|Z=z,X=x)&=\beta_0(x)\text{Var}(A|Z=z,X=x)+\rho_0(x)\\
&+\sum_{j=1,2}m_{0j}(z,x){\psi}_{0j}(x)+\sum_{k=0,1,2}m_{1k}(z,x)\psi_{1k}(x),
\end{aligned}
\end{equation}
holds under Assumptions \ref{assp:2} and $3^{\prime}$, where $\beta_0(x):=E\{\beta(X,U)|X=x\}$ and
\begin{equation*}
\begin{aligned}
\rho_0(x)&:=\text{Cov}\{\xi_y(U,X),\xi_a(U,X)|X=x\}\\
\psi_{01}(x)&:=\text{Cov}\{\beta(U,X),\xi_a(U,X)|X=x\};\\
\psi_{02}(x)&:=\text{Cov}\{\theta_1(U,X),\xi_a(U,X)|X=x\};\\
\psi_{10}(x)&:=\text{Cov}\{\xi_y(U,X),\theta_2(U,X)|X=x\};\\
\psi_{11}(x)&:=\text{Cov}\{\beta(U,X),\theta_2(U,X)|X=x\};\\
\psi_{12}(x)&:=\text{Cov}\{\theta_1(U,X),\theta_2(U,X)|X=x\}.
\end{aligned}
\end{equation*}
It is straightforward to verify that equation (\ref{eq:ett}) holds if for all $j=1,2$ and $k=0,1,2$
\begin{equation}
\begin{aligned}
\label{eq:ortho}
\psi_{0j}(X)=0 \text{ and } \psi_{1k}(X)=0,
\end{aligned}
\end{equation}
almost surely. The orthogonality condition (\ref{eq:ortho}) does not rule out non-linear forms of essential heterogeneity  \citep{2019arXiv170907779T}. In particular, if $\beta_0(X,U)=\beta_0(X)$ and $\theta_j(X,U)=\theta_j(X)$ for $j=1,2$ almost surely (i.e., Assumption \ref{assp:3} holds), then (\ref{eq:ortho}) holds. Assumption \ref{assp:4} ensures that (\ref{eq:estimand}) is well-defined while Assumption \ref{assp:1} imbues the identifying functional therein with causal interpretation as the ATE. The rest of the proof below follows from (\ref{eq:ett}).
\subsubsection*{Proof of explicit representation (i)}
\begin{eqnarray*}
\begin{aligned}
 E\{\varphi_1(O;\pi_0,\mu_0)|X\}=\frac{E(\varepsilon Y |Z=1,X)-E(\varepsilon Y |Z=0,X)}{\text{Var}(A|Z=1,X)-\text{Var}(A|Z=0,X)}=\beta_0(X).
 \end{aligned}
\end{eqnarray*}

\subsubsection*{Proof of implicit representation (ii)}
\begin{eqnarray*}
\begin{aligned}
E\{\varphi_2(O;\pi_0,\beta_0,\tau_0)|X\}&=E\left\{\frac{(2Z-1)\rho_0(X)}{\pi_0(Z|X)}\biggr\rvert X\right\}=0. 
 \end{aligned}
\end{eqnarray*}

\subsubsection*{Proof of implicit representation (iii)}
\begin{eqnarray*}
\begin{aligned}
E\{\varphi_3(O;\mu_0,\beta_0,\rho_0)|Z,X\}&=\rho_0(X)-\rho_0(X)=0.
 \end{aligned}
\end{eqnarray*}

  \subsection*{Proof of Theorem \ref{thr:efficient}}

We follow closely the semiparametric efficiency theory of \cite{newey1990semiparametric} and \cite{bickel1993efficient}. Consider a parametric submodel for the law of the observed data, $$f_t(o)=f_t(y|a,z,x)\mu_t(z,x)^a\{1-\mu_t(z,x)\}^{1-a}\pi_t(x)^z\{1-\pi_t(x)\}^{1-z}f_t(x),$$ where $\mu_t(z,x):= P_t(A=1|Z=z,X=x)$ and $\pi_t(x):= P_t(Z=1|X=x)$. The score function $S_t(o)$ is given by $ S_{t}(y|a,z,x) +S_{t}(a|z,x)+ S_{t}(z|x) + S_{t}(x),$ where $S_{t}(y|a,z,x)=\partial  \log f_t(y|a,z,x)/\partial t$, $S_{t}(a|z,x)=\frac{a-\mu_t(z,x)}{\mu_t(z,x)\{1-\mu_t(z,x)\}}\frac{\partial \mu_t(z,x)}{\partial t}$,  $S_{t}(z|x)=\frac{z-\pi_t(x)}{\pi_t(x)\{1-\pi_t(x)\}}\frac{\partial \pi_t(x)}{\partial t}$ and $S_t(x)=\log f_t(x)/\partial t$. A representation of the tangent space is therefore given by $$\mathcal{T}=\left\{S_{t}(y|a,z,x)+\{a-\mu_t(z,x)\}\varrho_{1,t}(z,x)+\{z-\pi_t(x)\}\varrho_{2,t}(x)+\varrho_{3,t}(x) \right\},$$ where $\{\varrho_{j,t}(\cdot)\}_{j=1,2,3}$ are arbitrary square-integrable functions. Pathwise differentiability follows if we can find a random element $G_t(O)\in \mathcal{T}$ such that it satisfies $E_{t}(G_t)=0$ and 
$\partial \gamma_t/\partial t=E_{t}\left\{G_t(O)S_{t}(O)\right\}$, where $E_t\{h(O)\}=\int h(o) dF_t$.  We make use of the following equalities in the proof, that for any arbitrary square-integrable functions $\varrho_1(A,Z,X)$, $\varrho_2(Z,X)$ and $\varrho_3(X)$,
\begin{eqnarray}
	\label{identity_1}	&& E_t\{\varrho_1(A,Z,X)S_{t}(Y|A,Z,X)\}  = 0; \\
	\label{identity_2}	 &&E_t\{\varrho_2(Z,X)S_{t}(A|Z,X)\}  =0; \quad E_t\left\{\varrho_2(X,Z)\varepsilon_t \right\}= 0; \\
	\label{identity_3}	 &&E_t\{\varrho_3(X)S_{t}(Z|X)\}  =0;\quad   E_t [\varrho_3(X)\{Z-\pi_{t}(Z|X)\}]= 0;\\
	\label{identity_4}	&&E_t\left\{\varphi_1(O;\pi_t,\mu_t)\rvert X\right\}=\beta_t(X),  
\end{eqnarray}
where (\ref{identity_4}) follows from the proof of Theorem \ref{thr:identification}. We start with representation (i) of Theorem \ref{thr:identification}. To ease notation, let $\sigma^2_t(z,x):=\mu_t(z,x)\{1-\mu_t(z,x)\}$ denote the conditional variance. Differentiating the right hand side of $\gamma_t=E_t\left\{\varphi_1(O;\pi_t,\mu_t)\right\}$ under the integral with respect to $t$ yields
\begin{eqnarray*}
\nabla _{t}\gamma _{t} &=&\nabla _{t}E_{t}\left\{ \frac{2Z-1 }{%
\pi_{t}(Z|X)}\frac{\varepsilon_t Y }{\left\{
\sigma _{t}^{2}(1,X)-\sigma _{t}^{2}(0,X)\right\} }\right\}  \\
&=&E_t\left\{\frac{2Z-1 }{%
\pi_{t}(Z|X)}\frac{\varepsilon_t YS_t\left( O\right) }{\left\{ \sigma _{t}^{2}(1,X)-\sigma _{t}^{2}(0,X)\right\} } \right\}  \\
&&-E_t\left\{ \frac{2Z-1 }{%
\pi_{t}(Z|X)}\frac{E_t\left( AS_t(A|Z,X)|Z,X\right) Y
}{\left\{\sigma _{t}^{2}(1,X)-\sigma _{t}^{2}(0,X)\right\} } \right\}  \\
&&-E_{t}\left\{ \frac{2Z-1 }{%
\pi_{t}(Z|X)}\frac{\varepsilon_t  Y\nabla _{t}\left\{ \sigma _{t}^{2}(1,X)-\sigma _{t}^{2}(0,X)\right\} }{\left\{\sigma _{t}^{2}(1,X)-\sigma _{t}^{2}(0,X)\right\} ^{2}} \right\}  \\
&&-E_{t}\left\{ \frac{2Z-1 }{%
\pi_{t}(Z|X)}\frac{\varepsilon_t Y S_{t}(Z|X)  }{\left\{ \sigma _{t}^{2}(1,X)-\sigma _{t}^{2}(0,X)\right\} } \right\}  \\
&:=& \mathscr{L}_1-\mathscr{L}_2-\mathscr{L}_3-\mathscr{L}_4.
\end{eqnarray*}

We consider the terms $\mathscr{L}_1$ to $\mathscr{L}_4$ separately:

\begingroup
\allowdisplaybreaks
 \begin{eqnarray*}
 \mathscr{L}_1 &=& E_t\left\{ \frac{2Z-1 }{%
\pi_{t}(Z|X)}\frac{\varepsilon_t YS_t\left( O\right) }{\left\{ \sigma _{t}^{2}(1,X)-\sigma _{t}^{2}(0,X)\right\} }%
 \right\};\\
\mathscr{L}_2 &=& E\left\{ \frac{2Z-1 }{%
\pi_{t}(Z|X)}\frac{E_t\left( AS_t(A|Z,X)|Z,X\right) Y
}{\left\{ \sigma _{t}^{2}(1,X)-\sigma _{t}^{2}(0,X)\right\} } \right\}  \\
&=& E\left\{\frac{2Z-1 }{%
\pi_{t}(Z|X)}\frac{E_t\left( AS_t(A|Z,X)|Z,X\right) E_t(Y|Z,X)
}{\left\{ \sigma _{t}^{2}(1,X)-\sigma _{t}^{2}(0,X)\right\} } \right\}\\
&=& E\left\{ \frac{2Z-1 }{%
\pi_{t}(Z|X)}\frac{ AS_t(A|Z,X) E_t(Y|Z,X)
}{\left\{  \sigma _{t}^{2}(1,X)-\sigma _{t}^{2}(0,X)\right\} } \right\}\\
&=& E\left\{ \frac{2Z-1 }{%
\pi_{t}(Z|X)}\frac{ \varepsilon_t S_t(A|Z,X) E_t(Y|Z,X)
}{\left\{ \sigma _{t}^{2}(1,X)-\sigma _{t}^{2}(0,X)\right\} } \right\} \quad \text{by (\ref{identity_2})}\\
&=& E\left\{\frac{2Z-1 }{%
\pi_{t}(Z|X)}\frac{ \varepsilon_t E_t(Y|Z,X) S_t(O)
}{\left\{ \sigma _{t}^{2}(1,X)-\sigma _{t}^{2}(0,X)\right\} }\right\}; \quad \text{by (\ref{identity_1}; \ref{identity_2})}\\
\mathscr{L}_3 &=&E_{t}\left\{ \frac{2Z-1 }{%
\pi_{t}(Z|X)}\frac{ \varepsilon_t  Y\nabla _{t}\left\{ \sigma
_{t}^{2}(1,X)-\sigma _{t}^{2}(0,X)\right\} }{\left\{ \sigma_t
^{2}(1,X)-\sigma_t^{2}(0,X)\right\} ^{2}} \right\}\\
&=&E_{t}\left\{\frac{2Z-1 }{%
\pi_{t}(Z|X)}\frac{ \varepsilon_t  Y[1-\mu_t(1,X)]E_t\left[AS_t(A|Z=1,X)|Z=1,X \right]  }{\left\{ \sigma
^{2}_t(1,X)-\sigma ^{2}_t(0,X)\right\} ^{2}} \right\}\\
&&-E_{t}\left\{\frac{2Z-1 }{%
\pi_{t}(Z|X)}\frac{ \varepsilon_t  Y\mu_t(1,X)E_t\left[AS_t(A|Z=1,X)|Z=1,X \right] }{\left\{ \sigma
^{2}_t(1,X)-\sigma ^{2}_t(0,X)\right\} ^{2}} \right\}\\
&&+E_{t}\left\{ \frac{2Z-1 }{%
\pi_{t}(Z|X)}\frac{ \varepsilon_t  Y[1-\mu_t(0,X)]E_t\left[AS_t(A|Z=0,X)|Z=0,X \right] }{\left\{ \sigma
^{2}_t(1,X)-\sigma ^{2}_t(0,X)\right\} ^{2}} \right\}\\
&&-E_{t}\left\{\frac{2Z-1 }{%
\pi_{t}(Z|X)}\frac{ \varepsilon_t Y\mu_t(0,X)E_t\left[AS_t(A|Z=0,X)|Z=0,X \right] }{\left\{ \sigma_t
^{2}(1,X)-\sigma_t^{2}(0,X)\right\} ^{2}} \right\}\\
&=&E_t\left\{ \frac{\beta_t\left( X\right) }{\left\{ \sigma_t
^{2}(1,X)-\sigma_t ^{2}(0,X)\right\} }\frac{Z}{\pi_t\left( Z|X\right) }\frac{%
 \varepsilon_t \sigma ^{2}_t(Z,X) S_t(O)  }{\mu_t(Z,X)}
\right\}  \\
&&-E_t\left\{ \frac{\beta_t\left( X\right) }{\left\{ \sigma
^{2}_t(1,X)-\sigma ^{2}_t(0,X)\right\} }\frac{Z}{\pi_t\left( Z|X\right) }\frac{%
 \varepsilon_t  \sigma ^{2}_t(Z,X) S_t(O)}{\left( 1-\mu_t(Z,X)\right) } \right\}  \\
&&-E_t\left\{ \frac{\beta_t\left( X\right) }{\left\{ \sigma_t
^{2}(1,X)-\sigma_t ^{2}(0,X)\right\} }\frac{1-Z}{\pi_{t}\left( Z|X\right) }\frac{%
 \varepsilon_t \sigma ^{2}_t(Z,X) S_t(O)}{\mu_t (Z,X)}
\right\}  \\
&&+E_t\left\{ \frac{\beta_t\left( X\right) }{\left\{ \sigma_t
^{2}(1,X)-\sigma_t ^{2}(0,X)\right\} }\frac{1-Z}{\pi_{t}\left( Z|X\right) }\frac{%
 \varepsilon_t  \sigma ^{2}_t(Z,X) S_t(O)}{\left[ 1-\mu_t (Z,X)\right] }\right\} \quad \text{by (\ref{identity_1}, \ref{identity_2}, \ref{identity_4})}\\
&=&E_t\left\{  \frac{2Z-1 }{%
\pi_{t}(Z|X)} \frac{\beta_t\left( X\right) }{\left\{ \sigma_t
^{2}(1,X)-\sigma_t ^{2}(0,X)\right\} }\frac{%
 \varepsilon_t  \sigma ^{2}_t(Z,X) S_t(O)}{\mu_t (Z,X)}
\right\}  \\
&&-E_t\left\{  \frac{2Z-1 }{%
\pi_{t}(Z|X)}\frac{\beta_t\left( X\right) }{\left\{ \sigma_t
^{2}(1,X)-\sigma_t ^{2}(0,X)\right\} }\frac{%
 \varepsilon_t \sigma ^{2}_t(Z,X) S_t(O)}{\left[ 1-\mu_t (Z,X)\right] }\right\};\\
  \mathscr{L}_4&=&E_t\left\{ \frac{2Z-1 }{\pi_t(Z|X)}\frac{\varepsilon_tYS_{t}(Z|X)   }{\left\{ \sigma
^{2}(1,X)-\sigma ^{2}(0,X)\right\} }  \right\} \\
&=&E_t\left\{ \left\{ 
\begin{array}{c}
E_t\left[ \frac{2Z-1 }{\pi_t(Z|X)}\frac{\varepsilon_t }{\left\{ \sigma _{t}^{2}(1,X)-\sigma
_{t}^{2}(0,X)\right\} }Y \biggr \rvert Z,X\right]  \\ 
-E_t\left[ \frac{2Z-1 }{\pi_t(Z|X)}\frac{\varepsilon_t}{\left\{ \sigma _{t}^{2}(1,X)-\sigma
_{t}^{2}(0,X)\right\} }Y \biggr \rvert X\right] 
\end{array}%
\right\} S_{t}(Z|X)\right\}  \quad \text{by (\ref{identity_3})}\\
&=&E_t\left\{ \left\{ 
\begin{array}{c}
E_t\left[ \frac{2Z-1}{\pi_t(Z|X)}\frac{\varepsilon_t}{\left\{ \sigma _{t}^{2}(1,X)-\sigma
_{t}^{2}(0,X)\right\} }Y \biggr \rvert Z,X\right]  \\ 
-E_t\left[ \frac{2Z-1 }{\pi_t(Z|X)}\frac{\varepsilon_t }{\left\{ \sigma _{t}^{2}(1,X)-\sigma
_{t}^{2}(0,X)\right\} }Y \biggr \rvert X\right] 
\end{array}%
\right\} S_{t}(O)\right\}   \quad \text{by (\ref{identity_1}, \ref{identity_2}, \ref{identity_3})}\\
&=&E_t\left\{ \left\{\frac{2Z-1 }{\pi_t(Z|X)}\left[\frac{\sigma _{t}^{2}(Z,X)\beta_t\left( X\right) }{\left\{
\sigma _{t}^{2}(1,X)-\sigma _{t}^{2}(0,X)\right\} }
+\frac{\rho_t(X)}{\left\{ \sigma _{t}^{2}(1,X)-\sigma
_{t}^{2}(0,X)\right\} }\right]-\beta_t(X)\right\}S_t(O)\right\},  
 \end{eqnarray*}
where the last equality holds from the proof of Theorem  \ref{thr:identification} as well as identity (\ref{identity_4}). Combining the terms $\mathscr{L}_1$ to $\mathscr{L}_4$ yields
\begin{eqnarray*}
\nabla _{t}\gamma_{t} &=&E\Biggr\{\Biggr[\frac{2Z-1 }{\pi_t(Z|X)}\frac{\varepsilon_t \left[ Y-\beta_{t}\left( X\right) \mu_t(Z,X) -\tau_t(Z,X) \right] }{\left\{ \sigma
^{2}_t(1,X)-\sigma ^{2}_t(0,X)\right\} } \\
&&-\frac{2Z-1 }{\pi_t(Z|X)}\frac{\sigma _{t}^{2}(Z,X)}{\left\{
\sigma _{t}^{2}(1,X)-\sigma _{t}^{2}(0,X)\right\} }\beta_{t}\left( X\right)
+\beta_{t}\left( X\right)  \\
&&-\frac{(2Z-1)\rho_t(X)}{\pi_t(Z|X)\left\{ \sigma _{t}^{2}(1,X)-\sigma
_{t}^{2}(0,X)\right\} }   \\
&&- \frac{2Z-1}{\pi_t(Z|X)} \frac{\varepsilon_t\beta_{t}\left( X\right)[1-\mu_t(Z,X)] }{\left\{ \sigma
^{2}_t(1,X)-\sigma ^{2}_t(0,X)\right\} }\\
&&+ \frac{2Z-1 }{\pi_t(Z|X)}\frac{\varepsilon_t \beta_{t}\left( X\right)\mu_t(Z,X)}{\left\{ \sigma
^{2}_t(1,X)-\sigma ^{2}_t(0,X)\right\} }\Biggr]S_t(O)\Biggr\}\\
&=&E\Biggr\{\Biggr[\frac{2Z-1 }{\pi_t(Z|X)}\frac{\varepsilon_t\left[ Y-\beta_{t}\left( X\right) A -\tau_t(Z,X) \right] }{\left\{ \sigma
^{2}_t(1,X)-\sigma ^{2}_t(0,X)\right\} } \\
&&+ \frac{2Z-1 }{\pi_t(Z|X)}\frac{\varepsilon_t\beta_{t}\left( X\right)A}{\left\{ \sigma
^{2}_t(1,X)-\sigma_t^{2}(0,X)\right\} }\\
&&-\frac{2Z-1 }{\pi_t(Z|X)}\frac{\sigma _{t}^{2}(Z,X)}{\left\{
\sigma _{t}^{2}(1,X)-\sigma _{t}^{2}(0,X)\right\} }\beta_{t}\left( X\right)
+\beta_{t}\left( X\right)  \\
&&-\frac{(2Z-1) \rho_t(X)}{\pi_t(Z|X)\left\{ \sigma _{t}^{2}(1,X)-\sigma
_{t}^{2}(0,X)\right\} }   \\
&&- \frac{2Z-1}{\pi_t(Z|X)} \frac{\varepsilon_t \beta_{t}\left( X\right)[1-\mu_t(Z,X)] }{\left\{ \sigma
^{2}_t(1,X)-\sigma ^{2}_t(0,X)\right\} }\Biggr]S_t(O)\Biggr\}\\
&=&E\Biggr\{\Biggr[\frac{2Z-1 }{\pi_t(Z|X)}\frac{\varepsilon_t \left[ Y-\beta_{t}\left( X\right) A -\tau_t(Z,X)\right] }{\left\{ \sigma
^{2}_t(1,X)-\sigma ^{2}_t(0,X)\right\} } \\
&&+ \frac{2Z-1 }{\pi_t(Z|X)}\frac{\sigma _{t}^{2}(Z,X)}{\left\{
\sigma _{t}^{2}(1,X)-\sigma _{t}^{2}(0,X)\right\} }\beta_{t}\left( X\right)\\
&&-\frac{2Z-1 }{\pi_t(Z|X)}\frac{\sigma _{t}^{2}(Z,X)}{\left\{
\sigma _{t}^{2}(1,X)-\sigma _{t}^{2}(0,X)\right\} }\beta_{t}\left( X\right)
+\beta_{t}\left( X\right)  \\
&&-\frac{(2Z-1) \rho_t(X)}{\pi_t(Z|X)\left\{ \sigma _{t}^{2}(1,X)-\sigma
_{t}^{2}(0,X)\right\} }  \Biggr]S_t(O)\Biggr\}\\
&=&E\Biggr\{\Biggr[\frac{2Z-1 }{\pi_t(Z|X)}\frac{\varepsilon_t \left[ Y-\beta_{t}\left( X\right) A -\tau_t(Z,X) \right] }{\left\{ \sigma
^{2}_t(1,X)-\sigma ^{2}_t(0,X)\right\} } \\
&&-\frac{(2Z-1)\rho_t(X)}{\pi_t(Z|X)\left\{ \sigma _{t}^{2}(1,X)-\sigma
_{t}^{2}(0,X)\right\} } +\beta_{t}\left( X\right)   \Biggr]S_t(O)\Biggr\}\\
&=&E\Biggr\{\Biggr[\frac{2Z-1}{\pi_t(Z|X)}\frac{\varepsilon_t \left[ Y-\beta_{t}\left( X\right) A -\tau_t(Z,X) \right] }{\left\{ \sigma
 _{t}^{2}(1,X)-\sigma  _{t}^{2}(0,X)\right\} } \\
&&-\frac{(2Z-1)\rho_t(X)}{\pi_t(Z|X)\left\{ \sigma _{t}^{2}(1,X)-\sigma
_{t}^{2}(0,X)\right\} }+\beta_{t}\left( X\right) -\gamma_t  \Biggr]S_t(O)\Biggr\}\\
&:=& E[\{\varphi_{\text{eff}}(O;\pi_t,\mu_t,\beta_t,\tau_t,\rho_t)-\gamma_t\}S_t(O)].
\end{eqnarray*}
We can readily verify that $\varphi_{\text{eff}}(O;\pi_t,\mu_t,\beta_t,\tau_t,\rho_t)-\gamma_t \in \mathcal{T}$. It follows that  $$\varphi_{\text{eff}}(O;\pi_0,\mu_0,\beta_0,\tau_0,\rho_0)-\gamma$$ is the efficient influence function for estimating $\gamma$ in $\mathcal{M}$ by Theorem 3.1 of  \cite{newey1990semiparametric}.

  \subsection*{Proof of Lemma \ref{lemma:if}}

It suffices to show that $\gamma=E\{\varphi_{\text{eff}}(O;{\pi}^{\ast}, {\mu}^{\ast},{\beta}^{\ast},{\tau}^{\ast},{\rho}^{\ast})\}=E\{\varphi_{\text{eff}}(O;{\eta}^{\ast})\}$ if at least one of the following holds: (i) Suppose $({\pi}^{\ast}, {\mu}^{\ast})=(\pi_0,\mu_0)$. Then
\begin{eqnarray*}
E\{\varphi_{\text{eff}}(O;{\eta}^{\ast})\}
&=&E\Biggr[
\frac{2Z-1}{\pi_0(Z|X)}\frac{\varepsilon\{Y-\beta_0(X)A\}}{\sigma ^{2}(1,X;\mu_0)-\sigma ^{2}(0,X;\mu_0)}\\
&&-\frac{2Z-1}{\pi_0(Z|X)}\frac{\varepsilon\{\beta^{\ast}(X)-\beta_0(X)\}A}{\sigma ^{2}(1,X;\mu_0)-\sigma ^{2}(0,X;\mu_0)}\\
&&-\frac{2Z-1}{\pi_0(Z|X)}\frac{\varepsilon{\tau}^{\ast}(Z,X)}{\sigma ^{2}(1,X;\mu_0)-\sigma ^{2}(0,X;\mu_0)}\\
&&-\frac{2Z-1}{\pi_0(Z|X)}\frac{{\rho}^{\ast}(X)}{\sigma ^{2}(1,X;\mu_0)-\sigma ^{2}(0,X;\mu_0)}+E\{\beta^{\ast}(X)\}\Biggr]\\
&=&E\Biggr[
\frac{2Z-1}{\pi_0(Z|X)}\frac{\rho_0(X)}{\sigma ^{2}(1,X;\mu_0)-\sigma ^{2}(0,X;\mu_0)}\\
&&-\frac{2Z-1}{\pi_0(Z|X)}\frac{\sigma^2(Z,X;\mu_0)\{\beta^{\ast}(X)-\beta_0(X)\}}{\sigma ^{2}(1,X;\mu_0)-\sigma ^{2}(0,X;\mu_0)}+\beta^{\ast}(X)\Biggr]\\
&=&E\Biggr[
\frac{2Z-1}{\pi_0(Z|X)}\frac{\rho_0(X)}{\sigma ^{2}(1,X;\mu_0)-\sigma ^{2}(0,X;\mu_0)}\\
&&-\{\beta^{\ast}(X)-\beta_0(X)\}+\beta^{\ast}(X)\Biggr]\\
&=&\gamma.
\end{eqnarray*}%

(ii) Next, suppose $({\pi}^{\ast},{\beta}^{\ast},{\tau}^{\ast})=(\pi_0,\beta_0,\tau_0)$. Then
\begin{eqnarray*}
E\{\varphi_{\text{eff}}(O;{\eta}^{\ast})\}
&=&E\Biggr[
\frac{2Z-1}{\pi_0(Z|X)}\frac{\varepsilon\{ Y-\beta_0(X)A\}}{\sigma ^{2}(1,X;\mu^{\ast})-\sigma ^{2}(0,X;\mu^{\ast})}\\
&&-\frac{2Z-1}{\pi_0(Z|X)}\frac{\{\mu^{\ast}(Z,X)-\mu_0( Z,X)\}\{ Y-\beta_0(X)A-\tau_0(Z,X)\}}{\sigma ^{2}(1,X;\mu^{\ast})-\sigma ^{2}(0,X;\mu^{\ast})}\\
&&-\frac{2Z-1}{\pi_0(Z|X)}\frac{\rho^{\ast}(X)}{\sigma ^{2}(1,X;\mu^{\ast})-\sigma ^{2}(0,X;\mu^{\ast})}+\beta_0(X)\Biggr]\\
&=&E\Biggr[
\frac{2Z-1}{\pi_0(Z|X)}\frac{\rho_0(X)}{\sigma ^{2}(1,X;\mu^{\ast})-\sigma ^{2}(0,X;\mu^{\ast})}+\beta_0(X)\Biggr]\\
&=&\gamma.
\end{eqnarray*}%

(iii) Finally, suppose $({\mu}^{\ast},{\beta}^{\ast}, {\rho}^{\ast})=(\mu_0,\beta_0,\rho_0)$. Then
\begin{eqnarray*}
E\{\varphi_{\text{eff}}(O;{\eta}^{\ast})\}
&=&E\Biggr[
\frac{2Z-1}{\pi^{\ast}(Z|X)}\frac{\varepsilon\{ Y-\beta_0(X)A\}-\rho_0(X)}{\sigma ^{2}(1,X;\mu_0)-\sigma ^{2}(0,X;\mu_0)}\\
&&-\frac{2Z-1}{\pi^{\ast}(Z|X)}\frac{\varepsilon\tau^{\ast}(Z,X)}{\sigma ^{2}(1,X;\mu_0)-\sigma ^{2}(0,X;\mu_0)}+\beta_0(X)\Biggr]\\
&=&E\Biggr[
\frac{2Z-1}{\pi^{\ast}(Z|X)}\frac{\rho_0(X)-\rho_0(X)}{\sigma ^{2}(1,X;\mu_0)-\sigma ^{2}(0,X;\mu_0)}+\beta_0(X)\Biggr]\\
&=&\gamma.
\end{eqnarray*}%

The last claim in Lemma  \ref{lemma:if} follows by noting that under the intersection submodel $\cap_{k=1}^3\mathcal{M}_{k}$,  $E\left\{{\partial  \varphi_{\text{eff}}(O;\eta)}/{\partial \eta^\T}\rvert_{\eta=\eta^{\ast}}\right\}=0$ by  Neyman orthogonality \citep{neyman1,neyman2,belloni2017program, 10.1111/ectj.12097, chernozhukov2020locally}.

\subsection*{Proof of Theorem \ref{thm:risk}}

To ease notation, let $n_{m}^s:= \#\{1\leq i\leq n: i\in I^s_{m}\}$ for $m=0,1$. By the definition of the proposed model selector,
\begin{align*}
&\max_{\substack{k\in\{1,2,3\}, {\alpha}^{\prime}_{+k}\in \mathscr{A}_k}} \frac{1}{S}\sum_{s=1}^S[\mathbb{P}^1_s\{\phi_s(\hat{\alpha}^{(1)}; \hat{\alpha}^{(1)}_{-k},\alpha^{\prime}_{+k})\}]^2\leq \max_{\substack{k\in\{1,2,3\}, {\alpha}^{\prime}_{+k}\in \mathscr{A}_k}} \frac{1}{S}\sum_{s=1}^S[\mathbb{P}^1_s\{\phi_s({\alpha}^{(1)}; {\alpha}^{(1)}_{-k},\alpha^{\prime}_{+k})\}]^2.
\end{align*}
It follows that
\begin{align*}
\widetilde{\mathcal{R}}^{(1)}({\hat{\alpha}^{(1)}}) \leq \max_{\substack{k\in\{1,2,3\}}}\frac{1}{S}\sum_{s=1}^S[\mathbb{P}^1_s\{\phi_s({\alpha}^{(1)}; {\alpha}^{(1)}_{-k},\dot{\alpha}_{+k})\}]^2,
\end{align*}
where $\dot{\alpha}_{+k}=\argmax_{{\alpha}^{\prime}_{+k}\in \mathscr{A}_k} \frac{1}{S}\sum_{s=1}^S[\mathbb{P}^1_s\{\phi_s({\alpha}^{(1)}; {\alpha}^{(1)}_{-k},{\alpha}^{\prime}_{+k})\}]^2$. By simple algebra, we have that for $k\in\{1,2,3\}$,
\begin{align*}
& [\PP^1_s \{\phi_s(\hat{\alpha}^{(1)}; \hat{\alpha}^{(1)}_{-k},\tilde{\alpha}_{+k})\}]^2=    \frac{1}{{n_1^s}^2} \sum_{i,j}  \Big[\PP^1  \{ \phi_s(\hat{\alpha}^{(1)}; \hat{\alpha}^{(1)}_{-k},\tilde{\alpha}_{+k})_i \} \PP^1 \{ \phi_s(\hat{\alpha}^{(1)}; \hat{\alpha}^{(1)}_{-k},\tilde{\alpha}_{+k})_j\}\Big] \\
 &+ \frac{2}{{n_1^s}^2} \sum_{i,j}  \Big [  \phi_s(\hat{\alpha}^{(1)}; \hat{\alpha}^{(1)}_{-k},\tilde{\alpha}_{+k})_i - \PP^1  \{ \phi_s(\hat{\alpha}^{(1)}; \hat{\alpha}^{(1)}_{-k},\tilde{\alpha}_{+k})_i\} \Big] \PP^1 \{  \phi_s(\hat{\alpha}^{(1)}; \hat{\alpha}^{(1)}_{-k},\tilde{\alpha}_{+k})_j \}\\
 &+ \frac{1}{{n_1^s}^2} \sum_{i,j} \Big [\phi_s(\hat{\alpha}^{(1)}; \hat{\alpha}^{(1)}_{-k},\tilde{\alpha}_{+k})_i - \PP^1  \{ \phi_s(\hat{\alpha}^{(1)}; \hat{\alpha}^{(1)}_{-k},\tilde{\alpha}_{+k})_i\} \Big]\times\\
&\phantom{+}\Big [\phi_s(\hat{\alpha}^{(1)}; \hat{\alpha}^{(1)}_{-k},\tilde{\alpha}_{+k})_j- \PP^1  \{\phi_s(\hat{\alpha}^{(1)}; \hat{\alpha}^{(1)}_{-k},\tilde{\alpha}_{+k})_j\} \Big],
\end{align*}
where $\phi_s(\hat{\alpha}^{(1)}; \hat{\alpha}^{(1)}_{-k},\tilde{\alpha}_{+k})_i$ denotes the estimating equation evaluated at $i$-th observation. Thus,
\begin{align*}
& [\PP^1_s \{\phi_s(\hat{\alpha}^{(1)}; \hat{\alpha}^{(1)}_{-k},\tilde{\alpha}_{+k})\}]^2=  [\PP^1 \{\phi_s(\hat{\alpha}^{(1)}; \hat{\alpha}^{(1)}_{-k},\tilde{\alpha}_{+k})\}]^2\\
 &+ \frac{2}{{n_1^s}^2} \sum_{i,j}  \Big [ \phi_s(\hat{\alpha}^{(1)}; \hat{\alpha}^{(1)}_{-k},\tilde{\alpha}_{+k})_i - \PP^1  \{\phi_s(\hat{\alpha}^{(1)}; \hat{\alpha}^{(1)}_{-k},\tilde{\alpha}_{+k})_i\} \Big] \PP^1 \{ \phi_s(\hat{\alpha}^{(1)}; \hat{\alpha}^{(1)}_{-k},\tilde{\alpha}_{+k})_j \}\\
& + \frac{1}{{n_1^s}^2} \sum_{i,j} \Big [\phi_s(\hat{\alpha}^{(1)}; \hat{\alpha}^{(1)}_{-k},\tilde{\alpha}_{+k})_i - \PP^1  \{ \phi_s(\hat{\alpha}^{(1)}; \hat{\alpha}^{(1)}_{-k},\tilde{\alpha}_{+k})_i\} \Big]\times\\
&\phantom{+}\Big[\phi_s(\hat{\alpha}^{(1)}; \hat{\alpha}^{(1)}_{-k},\tilde{\alpha}_{+k})_j- \PP^1  \{\phi_s(\hat{\alpha}^{(1)}; \hat{\alpha}^{(1)}_{-k},\tilde{\alpha}_{+k})_j\} \Big].
\end{align*}
The same decomposition holds for $[\PP^1_s \{\phi_s({\alpha}^{(1)}; {\alpha}^{(1)}_{-k},\dot{\alpha}_{+k})\}]^2$. By definition of our estimator, for any $\epsilon>0$, we have that
\begin{align*}
&\widetilde{\mathcal{R}}^{(1)}({\hat{\alpha}^{(1)}}) \leq  (1+2\epsilon)\max_{\substack{k}} \frac{1}{S}\sum_{s=1}^S  [\PP^1 \{\phi_s({\alpha}^{(1)}; {\alpha}^{(1)}_{-k},\dot{\alpha}_{+k})\}]^2\\
+& \Big\{ (1+\epsilon)\max_{\substack{k}} \frac{1}{S}\sum_{s=1}^S(   [\PP^1_s \{\phi_s({\alpha}^{(1)}; {\alpha}^{(1)}_{-k},\dot{\alpha}_{+k})\}]^2-  [\PP^1 \{\phi_s({\alpha}^{(1)}; {\alpha}^{(1)}_{-k},\dot{\alpha}_{+k})\}]^2 ) \\
&-\epsilon\max_{\substack{k}} \frac{1}{S}\sum_{s=1}^S [\PP^1\{\phi_s(\hat{\alpha}^{(1)}; \hat{\alpha}^{(1)}_{-k},\tilde{\alpha}_{+k})\}]^2   \Big\}\\
-& \Big\{ (1+\epsilon)\max_{\substack{k}} \frac{1}{S}\sum_{s=1}^S (   [\PP^1_s \{\phi_s(\hat{\alpha}^{(1)}; \hat{\alpha}^{(1)}_{-k},\tilde{\alpha}_{+k})\}]^2-  [\PP^1 \{\phi_s(\hat{\alpha}^{(1)}; \hat{\alpha}^{(1)}_{-k},\tilde{\alpha}_{+k})\}]^2 ) \\
& + \epsilon\max_{\substack{k}} \frac{1}{S}\sum_{s=1}^S [\PP^1\{\phi_s(\hat{\alpha}^{(1)}; \hat{\alpha}^{(1)}_{-k},\tilde{\alpha}_{+k})\}]^2    \Big\}.
\end{align*}
Combined with the decomposition of $[\PP^1_s \{\phi_s(\hat{\alpha}^{(1)}; \hat{\alpha}^{(1)}_{-k},\tilde{\alpha}_{+k})\}]^2$ and $[\PP^1_s \{\phi_s({\alpha}^{(1)}; {\alpha}^{(1)}_{-k},\dot{\alpha}_{+k})\}]^2$, we further have that
\begin{align*}
&\widetilde{\mathcal{R}}^{(1)}({\hat{\alpha}^{(1)}}) \leq (1+2\epsilon)\max_{\substack{k}} \frac{1}{S} \sum_{s=1}^S  [\PP^1 \{\phi_s({\alpha}^{(1)}; {\alpha}^{(1)}_{-k},\dot{\alpha}_{+k})\}]^2\\
+ & \Big\{ (1+\epsilon)  \max_{\substack{k}} \frac{1}{S}\sum_{s=1}^S \Big[ \frac{2}{n_1^s} \sum_{i}  ( \phi_s({\alpha}^{(1)}; {\alpha}^{(1)}_{-k},\dot{\alpha}_{+k})_i- \PP^1 \{\phi_s({\alpha}^{(1)}; {\alpha}^{(1)}_{-k},\dot{\alpha}_{+k})_i\})\PP^1 \{\phi_s({\alpha}^{(1)}; {\alpha}^{(1)}_{-k},\dot{\alpha}_{+k})\} \\
 + & \frac{1}{({n_1^s})^2}  \sum_{i,j}  \Big ( \phi_s({\alpha}^{(1)}; {\alpha}^{(1)}_{-k},\dot{\alpha}_{+k})_i - \PP^1  \{  \phi_s({\alpha}^{(1)}; {\alpha}^{(1)}_{-k},\dot{\alpha}_{+k})_i \} \Big)\times\\
 &\Big ( \phi_s({\alpha}^{(1)}; {\alpha}^{(1)}_{-k},\dot{\alpha}_{+k})_j - \PP^1  \{  \phi_s({\alpha}^{(1)}; {\alpha}^{(1)}_{-k},\dot{\alpha}_{+k})_j \} \Big)\Big] -\epsilon    \max_{\substack{k}} \frac{1}{S}\sum_{s=1}^S [\PP^1 \{\phi_s({\alpha}^{(1)}; {\alpha}^{(1)}_{-k},\dot{\alpha}_{+k})\}]^2  \Big\}\\
 - &\Big\{ (1+\epsilon) \max_{\substack{k}} \frac{1}{S}\sum_{s=1}^S \Big[  \frac{2}{n_1^s}  \sum_{i}  ( \phi_s(\hat{\alpha}^{(1)}; \hat{\alpha}^{(1)}_{-k},\tilde{\alpha}_{+k})_i- \PP^1 \{\phi_s(\hat{\alpha}^{(1)}; \hat{\alpha}^{(1)}_{-k},\tilde{\alpha}_{+k})_i\})\PP^1 \{\phi_s(\hat{\alpha}^{(1)}; \hat{\alpha}^{(1)}_{-k},\tilde{\alpha}_{+k})\}\\
  + & \frac{1}{({n_1^s})^2}  \sum_{i,j}  \Big (\phi_s(\hat{\alpha}^{(1)}; \hat{\alpha}^{(1)}_{-k},\tilde{\alpha}_{+k})_i - \PP^1  \{  \phi_s(\hat{\alpha}^{(1)}; \hat{\alpha}^{(1)}_{-k},\tilde{\alpha}_{+k})_i \} \Big) \times\\
  &\Big (\phi_s(\hat{\alpha}^{(1)}; \hat{\alpha}^{(1)}_{-k},\tilde{\alpha}_{+k})_j - \PP^1  \{  \phi_s(\hat{\alpha}^{(1)}; \hat{\alpha}^{(1)}_{-k},\tilde{\alpha}_{+k})_j \} \Big)\Big] + \epsilon  \max_{\substack{k}} \frac{1}{S}\sum_{s=1}^S  [\PP^1 \{\phi_s(\hat{\alpha}^{(1)}; \hat{\alpha}^{(1)}_{-k},\tilde{\alpha}_{+k})\}]^2   \Big\}.
 \end{align*}
Note that the only assumption on $\{I_s^m\}_{m=0,1}$ is its stochastic independence of the observations, we omit sup-index $s$ hereinafter.
Because the maximum of sum is at most the sum of maxima, we deal with the first order and second order terms separately.
By Lemma 2.2 in \cite{van2006oracle}, we further have the following bounds for the first order term,
\begin{align*}
& \PP^0 \Biggr[\max_{\substack{k\in\{1,2,3\}, {\alpha}, {\alpha}^{\prime}_{+k}\in \mathscr{A}_k}}  \Biggr\{ \frac{2(1+\epsilon)\sqrt \nn}{\nn} \sum_{i}  \Big ( \phi({\alpha}; {\alpha}_{-k},{\alpha}^{\prime}_{+k})_i - \PP^1   \big\{\phi({\alpha}; {\alpha}_{-k},{\alpha}^{\prime}_{+k})_i \big\}  \Big)  \\
&\times \PP^1   \big\{ \phi({\alpha}; {\alpha}_{-k},{\alpha}^{\prime}_{+k})\bigr\}- \epsilon \sqrt \nn  [\PP^1   \big\{\phi({\alpha}; {\alpha}_{-k},{\alpha}^{\prime}_{+k})\big\} ]^2 \Biggr\}\Biggr] \\
\leq &\PP^0  \Biggr(  \frac{16(1+ \epsilon)}{n_1^{1/q-1/2}} \log\{1+(r_3r_4r_5)^2(r_2r_5)^2(r_1r_4)^2 \}\times\\
&\max_{\substack{k\in\{1,2,3\}, {\alpha}, {\alpha}^{\prime}_{+k}\in \mathscr{A}_k}}   \bigg [\frac{||  \phi({\alpha}; {\alpha}_{-k},{\alpha}^{\prime}_{+k})\PP^1   \{\phi({\alpha}; {\alpha}_{-k},{\alpha}^{\prime}_{+k})\} ||_\infty}{\nn^{1-1/q}} \\
+& \bigg(\frac{3\PP^1  [\phi({\alpha}; {\alpha}_{-k},{\alpha}^{\prime}_{+k}) \PP^1   \big\{ \phi({\alpha}; {\alpha}_{-k},{\alpha}^{\prime}_{+k})\big\} ]^2 2^{(1-q)}(1+\delta)^{(2-q)}}{\epsilon^{2-q} [ \PP^1   \big\{ \phi({\alpha}; {\alpha}_{-k},{\alpha}^{\prime}_{+k})\big\}]^{2-2q}} \bigg)^{1/q} \bigg ]  \Biggr):= (I),
\end{align*}
and
\begin{align*}
& \PP^0\Biggr[\max_{\substack{k\in\{1,2,3\}, {\alpha}, {\alpha}^{\prime}_{+k}\in \mathscr{A}_k}}    \Biggr\{-\Big[ \frac{2(1+\epsilon)\sqrt \nn}{\nn} \sum_{i}  \Big ( \phi({\alpha}; {\alpha}_{-k},{\alpha}^{\prime}_{+k})_i - \PP^1   \big\{ \phi({\alpha}; {\alpha}_{-k},{\alpha}^{\prime}_{+k})_i\big\}  \Big) \times\\
& \PP^1   \big\{\phi({\alpha}; {\alpha}_{-k},{\alpha}^{\prime}_{+k})\big\}+ \epsilon \sqrt \nn  [\PP^1   \big\{ \phi({\alpha}; {\alpha}_{-k},{\alpha}^{\prime}_{+k})\big\} ]^2 \Biggr\}\Biggr]  \leq (I),
\end{align*}
where the maximum is taken over sets that fixes ${\alpha}_{-k}$ as ${\alpha}$ and varies ${\alpha}^{\prime}_{+k}$, for $k\in\{1,2,3\}$. Thus, $ \mathbb{P}^0\{\widetilde{\mathcal{R}}^{(1)}({\hat{\alpha}^{(1)}})\}$ is further bounded by
\begin{align*}
& \mathbb{P}^0\{\widetilde{\mathcal{R}}^{(1)}({\hat{\alpha}^{(1)}})\}\\
&\leq (1+2\epsilon)\PP^0\Big(\max_{\substack{k}}[\PP^1\{\phi({\alpha}^{(1)}; {\alpha}^{(1)}_{-k},\dot{\alpha}_{+k})\}]^2\Big) +  \frac{2}{\sqrt \nn}(I)\\
 + & \mathbb{P}^0\biggr[\max_{\substack{k}}\Big \{ \frac{1+\epsilon}{n_1^{2}} \sum_{i,j}  \Big ( \phi({\alpha}^{(1)}; {\alpha}^{(1)}_{-k},\dot{\alpha}_{+k})_i - \PP^1  \{ \phi({\alpha}^{(1)}; {\alpha}^{(1)}_{-k},\dot{\alpha}_{+k})_i\} \Big)\times\\
 &\Big (\phi({\alpha}^{(1)}; {\alpha}^{(1)}_{-k},\dot{\alpha}_{+k})_j - \PP^1  \{ \phi({\alpha}^{(1)}; {\alpha}^{(1)}_{-k},\dot{\alpha}_{+k})_j\} \Big) \\
  -& \frac{1+\epsilon}{n_1^{2}}  \sum_{i,j}  \Big (\phi(\hat{\alpha}^{(1)}; \hat{\alpha}^{(1)}_{-k},\tilde{\alpha}_{+k})_i- \PP^1  \{  \phi(\hat{\alpha}^{(1)}; \hat{\alpha}^{(1)}_{-k},\tilde{\alpha}_{+k})_i\} \Big)\times\\
  & \Big (\phi(\hat{\alpha}^{(1)}; \hat{\alpha}^{(1)}_{-k},\tilde{\alpha}_{+k})_j - \PP^1  \{ \phi(\hat{\alpha}^{(1)}; \hat{\alpha}^{(1)}_{-k},\tilde{\alpha}_{+k})_j\} \Big)\Big \}\biggr].
\end{align*}
Following Lemmas 4 and 5 of \cite{cui2019biasaware}, the U-statistics are bounded and we have the following excess risk bound,	
\begin{align*}	
& \mathbb{P}^0\{\widetilde{\mathcal{R}}^{(1)}({\hat{\alpha}^{(1)}})\}\\
&\leq (1+2\epsilon)\PP^0\Big(\max_{\substack{k}}[\PP^1\{\phi({\alpha}^{(1)}; {\alpha}^{(1)}_{-k},\dot{\alpha}_{+k})\}]^2\Big) \\
+ & (1+\epsilon)C \Bigg\{ \left( \frac{2M}{n_1^2} \log\left(1+\frac{M (r_3r_4r_5)^2(r_2r_5)^2(r_1r_4)^2 }{2}\right ) \right)^{1/2} \\
&+ \frac{2M}{n_1} \log\left(1+\frac{M (r_3r_4r_5)^2(r_2r_5)^2(r_1r_4)^2 }{2}\right ) \\
&+  \frac{4M^{3/2}}{n_1^{3/2}} \log^{3/2}\left(1+\frac{M (r_3r_4r_5)^2(r_2r_5)^2(r_1r_4)^2 }{2}+D_0\right )+\\
& \frac{4 M^2}{n_1^2} \log^{2}\left(1+\frac{M (r_3r_4r_5)^2(r_2r_5)^2(r_1r_4)^2 }{2}+D_1\right ) \Bigg\}\\
+ &  \PP^0  \Biggr(  \frac{16(1+ \epsilon)}{n_1^{1/q-1/2}} \log\{1+(r_3r_4r_5)^2(r_2r_5)^2(r_1r_4)^2 \}\times\\
&\max_{\substack{k\in\{1,2,3\}, {\alpha}, {\alpha}^{\prime}_{+k}\in \mathscr{A}_k}}   \bigg [\frac{||  \phi({\alpha}; {\alpha}_{-k},{\alpha}^{\prime}_{+k})\PP^1   \{\phi({\alpha}; {\alpha}_{-k},{\alpha}^{\prime}_{+k})\} ||_\infty}{\nn^{1-1/q}} \\
+& \bigg(\frac{3\PP^1  [\phi({\alpha}; {\alpha}_{-k},{\alpha}^{\prime}_{+k}) \PP^1   \big\{ \phi({\alpha}; {\alpha}_{-k},{\alpha}^{\prime}_{+k})\big\} ]^2 2^{(1-q)}(1+\delta)^{(2-q)}}{\epsilon^{2-q} [ \PP^1   \big\{ \phi({\alpha}; {\alpha}_{-k},{\alpha}^{\prime}_{+k})\big\}]^{2-2q}} \bigg)^{1/q} \bigg ]  \Biggr),
 \end{align*}	
where $C$, $M$, $D_0$, and $D_1$ are some universal constants. Finally, recall that for the term $$(1+2\epsilon)\PP^0\Big(\max_{\substack{k}}[\PP^1\{\phi({\alpha}^{(1)}; {\alpha}^{(1)}_{-k},\dot{\alpha}_{+k})\}]^2\Big),$$ $\dot{\alpha}_{+k}$ is  chosen corresponding to ${\alpha}^{(1)}$ under measure $\PP_s^1$. It is further bounded by  $(1+2\epsilon)\mathbb{P}^0\{\bar{\mathcal{R}}^{(1)}({{\alpha}^{(1)}})\}$, where $\bar{\alpha}_{+k}$ is chosen corresponding to ${\alpha}^{(1)}$ under true measure $\PP^1$. This completes the proof for the risk bound of the minimax estimator $\hat{\alpha}^{(1)}$.  Although details are omitted, the proof for the risk bound of the mixed minimax estimator $\hat{\alpha}^{(2)}$ is essentially the same. 	
  
      \endgroup
\vskip 0.2in

\section*{Appendix B.}
For binary treatment, the baseline risk $\xi_a(X,U)=p_0(X,U)$ and the risk difference ${\theta}_2(X)=p_1(X,U)-p_0(X,U)$ are variation dependent, where $p_z(X,U):=P(A=1|Z=z,X,U)$. \cite{ richardson2017modeling} showed that the log odds product
$$\omega_1(X,U):=\log\left[\frac{p_0(X,U)p_1(X,U)}{\{1-p_0(X,U)\}\{1-p_1(X,U)\}}\right],$$ is variation independent of $\omega_2(X):=\text{arctanh}\{{\theta}_2(X)\}$, which yields the appropriate linear structural model  
$P(A=1|Z,X,U;\omega_1,\omega_2)=\theta_2(X;\omega_2)Z+\xi_a(X,U; \omega_1,\omega_2)$ for binary treatment in (\ref{eq:semiparametric2}), where
\begin{align*}
\theta_2(X;\omega_2)&=\text{tanh}\{{\omega}_2(X)\};\\ 
\xi_a(X,U; \omega_1,\omega_2)&=\frac{e^{\omega_1}(2-\theta_2)+\theta_2-\sqrt{\{e^{\omega_1}(\theta_2-2)-\theta_2\}^2+4e^{\omega_1}(1-\theta_2)(1-e^{\omega_1})}}{2(e^{\omega_1}-1)}.
\end{align*}

\newpage
\section*{Appendix C.}

\newcommand\mycommfont[1]{\footnotesize\ttfamily\textcolor{blue}{#1}}
\SetCommentSty{mycommfont}

\SetKwInput{KwInput}{Input}                
\SetKwInput{KwOutput}{Output}              
\let\oldnl\nl
\newcommand{\nonl}{\renewcommand{\nl}{\let\nl\oldnl}}

\begin{algorithm}[!htbp]
  \caption{Selective machine learning algorithm to estimate the average treatment effect with an invalid instrumental variable}  \label{alg:sml}
  \KwInput{Dataset $\mathcal O=\{O_1,...,O_n\}$ and $\prod_{j=1}^5 r_j$ candidate learners}
  \For{$s = 1$ \textbf{to} $S$}
{\ShowLn In the training dataset $\mathcal{O}^{0s}$:\\   Construct the estimators $\hat{\eta}(\alpha;s)$ for all $\alpha \in \mathscr{A}:=\{(\alpha_1,\alpha_2,\alpha_3,\alpha_4,\alpha_5):1\leq \alpha_j\leq r_j \text{ for }j=1,...,5\}$\;

\ShowLn In the validation dataset $\mathcal O^{1s}$:\\ 
\ForEach{$\alpha^{\ast}\in \mathscr{A}$} {Evaluate $\mathbb{P}^1_s\{\phi_s(\alpha^{\ast};\alpha^{\ast}_{-k},\alpha_{+k})\}$ and $\mathbb{P}^1_s\{\phi_s(\alpha^{\ast}_{-k},\alpha^{\prime}_{+k};\alpha^{\ast}_{-k},\alpha_{+k})\}$, 
for all $\alpha_{+k},\alpha^{\prime}_{+k}\in \mathscr{A}_{k}:=\{(\alpha_j)_{j\in \mathscr{C}_k}:1\leq {\alpha}_j\leq r_j\}$ and $k\in\{1,2,3\}$; }
  }
\ForEach{$\alpha^{\ast}\in \mathscr{A}$} {\ShowLn Average the perturbations over the $S$ splits to obtain 
$$\hat{\Delta}^{(1)}(\alpha^{\ast};\alpha_{+k}):=\frac{1}{S}\sum_{s=1}^S[\mathbb{P}^1_s\{\phi_s(\alpha^{\ast};\alpha^{\ast}_{-k},\alpha_{+k})\}]^2$$
and $$\hat{\Delta}^{(2)}(\alpha^{\ast};\alpha_{+k},\alpha^{\prime}_{+k}):=\frac{1}{S}\sum_{s=1}^S[\mathbb{P}^1_s\{\phi_s(\alpha^{\ast}_{-k},\alpha^{\prime}_{+k};\alpha^{\ast}_{-k},\alpha_{+k})\}]^2,$$
for all $\alpha_{+k},\alpha^{\prime}_{+k}\in \mathscr{A}_{k}:=\{(\alpha_j)_{j\in \mathscr{C}_k}:1\leq {\alpha}_j\leq r_j\}$ and $k\in\{1,2,3\}$;

\ShowLn Evaluate $$\widehat{\Lambda}^{(1)}_k({\alpha^{\ast}})=\max_{\substack{\alpha_{+k} \in\mathscr{A}_{k}}}\hat{\Delta}^{(1)}(\alpha^{\ast};\alpha_{+k})$$ and $$\widehat{\Lambda}^{(2)}_k({\alpha^{\ast}})=\max_{\substack{\alpha_{+k},\alpha^{\prime}_{+k} \in\mathscr{A}_{k}}}\hat{\Delta}^{(2)}(\alpha^{\ast};\alpha_{+k},\alpha^{\prime}_{+k}),$$ for all $k\in\{1,2,3\}$; 

\ShowLn Evaluate the empirical pseudo-risks  $\widehat{\mathcal{R}}^{(1)}(\alpha^{\ast})=\max_{k\in\{1,2,3\}}\widehat{\Lambda}^{(1)}_k(\alpha^{\ast})$ and $\widehat{\mathcal{R}}^{(2)}(\alpha^{\ast})=\sum_{k=1}^3 \widehat{\Lambda}^{(2)}_k(\alpha^{\ast})$;\\}
 
\ShowLn Select the minimizers $\hat{\alpha}^{(1)}=\argmin_{\substack{{\alpha} }}\widehat{\mathcal{R}}^{(1)}({\alpha})$ and $\hat{\alpha}^{(2)}=\argmin_{\substack{{\alpha} }}\widehat{\mathcal{R}}^{(2)}({\alpha})$ as our nuisance parameter learners, and obtain the selective machine learning estimators $\hat{\gamma}^{(1)}_{sml}=\frac{1}{S}\sum_{s=1}^S\mathbb{P}^1_s\{{\varphi}_{\text{eff}}(\hat{\eta}(\hat{\alpha}^{(1)};s))\}$ and $\hat{\gamma}^{(2)}_{sml}=\frac{1}{S}\sum_{s=1}^S\mathbb{P}^1_s\{{\varphi}_{\text{eff}}(\hat{\eta}(\hat{\alpha}^{(2)};s))\}$
 
  \KwRet{$\left(\hat{\alpha}^{(1)},\hat{\gamma}^{(1)}_{sml}\right)$, $\left(\hat{\alpha}^{(2)},\hat{\gamma}^{(2)}_{sml}\right)$}
 
\end{algorithm}

\bibliography{bibliography}

\begin{thebibliography}{103}
\providecommand{\natexlab}[1]{#1}
\providecommand{\url}[1]{\texttt{#1}}
\expandafter\ifx\csname urlstyle\endcsname\relax
  \providecommand{\doi}[1]{doi: #1}\else
  \providecommand{\doi}{doi: \begingroup \urlstyle{rm}\Url}\fi

\bibitem[Abadie(2003)]{abadie2003semiparametric}
Alberto Abadie.
\newblock Semiparametric instrumental variable estimation of treatment response
  models.
\newblock \emph{Journal of Econometrics}, 113\penalty0 (2):\penalty0 231--263,
  2003.

\bibitem[Angrist et~al.(1996)Angrist, Imbens, and
  Rubin]{angrist1996identification}
Joshua~D Angrist, Guido~W Imbens, and Donald~B Rubin.
\newblock Identification of causal effects using instrumental variables.
\newblock \emph{Journal of the American statistical Association}, 91\penalty0
  (434):\penalty0 444--455, 1996.

\bibitem[Baiocchi et~al.(2014)Baiocchi, Cheng, and
  Small]{baiocchi2014instrumental}
Michael Baiocchi, Jing Cheng, and Dylan~S Small.
\newblock Instrumental variable methods for causal inference.
\newblock \emph{Statistics in Medicine}, 33\penalty0 (13):\penalty0 2297--2340,
  2014.

\bibitem[Bang and Robins(2005)]{bang2005doubly}
Heejung Bang and James~M Robins.
\newblock Doubly robust estimation in missing data and causal inference models.
\newblock \emph{Biometrics}, 61\penalty0 (4):\penalty0 962--973, 2005.

\bibitem[Belloni et~al.(2017)Belloni, Chernozhukov, Fern{\'a}ndez-Val, and
  Hansen]{belloni2017program}
Alexandre Belloni, Victor Chernozhukov, Ivan Fern{\'a}ndez-Val, and Christian
  Hansen.
\newblock Program evaluation and causal inference with high-dimensional data.
\newblock \emph{Econometrica}, 85\penalty0 (1):\penalty0 233--298, 2017.

\bibitem[Benjamin(2003)]{benjamin2003does}
Daniel~J Benjamin.
\newblock Does 401 (k) eligibility increase saving?: Evidence from propensity
  score subclassification.
\newblock \emph{Journal of Public Economics}, 87\penalty0 (5-6):\penalty0
  1259--1290, 2003.

\bibitem[Bickel et~al.(1993)Bickel, Klaassen, Bickel, Ritov, Klaassen, Wellner,
  and Ritov]{bickel1993efficient}
Peter~J Bickel, Chris~AJ Klaassen, Peter~J Bickel, Y~Ritov, J~Klaassen, Jon~A
  Wellner, and YA'Acov Ritov.
\newblock \emph{{Efficient and Adaptive Estimation for Semiparametric Models}}.
\newblock Johns Hopkins University Press Baltimore, 1993.

\bibitem[Bowden et~al.(2016)Bowden, Davey~Smith, Haycock, and
  Burgess]{bowden2016consistent}
Jack Bowden, George Davey~Smith, Philip~C Haycock, and Stephen Burgess.
\newblock Consistent estimation in {Mendelian} randomization with some invalid
  instruments using a weighted median estimator.
\newblock \emph{Genetic Epidemiology}, 40\penalty0 (4):\penalty0 304--314,
  2016.

\bibitem[Bowden and Turkington(1990)]{bowden1990instrumental}
Roger~J Bowden and Darrell~A Turkington.
\newblock \emph{Instrumental Variables}, volume~8.
\newblock Cambridge University Press, 1990.

\bibitem[Breiman(2001)]{breiman2001random}
Leo Breiman.
\newblock Random forests.
\newblock \emph{Machine Learning}, 45\penalty0 (1):\penalty0 5--32, 2001.

\bibitem[Chan(2013)]{chan2013simple}
Kwun Chuen~Gary Chan.
\newblock A simple multiply robust estimator for missing response problem.
\newblock \emph{Stat}, 2\penalty0 (1):\penalty0 143--149, 2013.

\bibitem[Chan et~al.(2014)Chan, Yam, et~al.]{chan2014oracle}
Kwun Chuen~Gary Chan, Sheung Chi~Phillip Yam, et~al.
\newblock Oracle, multiple robust and multipurpose calibration in a missing
  response problem.
\newblock \emph{Statistical Science}, 29\penalty0 (3):\penalty0 380--396, 2014.

\bibitem[Chen and Haziza(2017)]{chen2017multiply}
Sixia Chen and David Haziza.
\newblock Multiply robust imputation procedures for the treatment of item
  nonresponse in surveys.
\newblock \emph{Biometrika}, 104\penalty0 (2):\penalty0 439--453, 2017.

\bibitem[Chernozhukov and Hansen(2004)]{doi:10.1162/0034653041811734}
Victor Chernozhukov and Christian Hansen.
\newblock The effects of 401(k) participation on the wealth distribution: An
  instrumental quantile regression analysis.
\newblock \emph{The Review of Economics and Statistics}, 86\penalty0
  (3):\penalty0 735--751, 2004.

\bibitem[Chernozhukov et~al.(2018)Chernozhukov, Chetverikov, Demirer, Duflo,
  Hansen, Newey, and Robins]{10.1111/ectj.12097}
Victor Chernozhukov, Denis Chetverikov, Mert Demirer, Esther Duflo, Christian
  Hansen, Whitney Newey, and James Robins.
\newblock {Double/debiased machine learning for treatment and structural
  parameters}.
\newblock \emph{The Econometrics Journal}, 21\penalty0 (1):\penalty0 C1--C68,
  01 2018.

\bibitem[Chernozhukov et~al.(2022)Chernozhukov, Escanciano, Ichimura, Newey,
  and Robins]{chernozhukov2020locally}
Victor Chernozhukov, Juan~Carlos Escanciano, Hidehiko Ichimura, Whitney~K.
  Newey, and James~M. Robins.
\newblock Locally robust semiparametric estimation.
\newblock \emph{Econometrica}, 90\penalty0 (4):\penalty0 1501--1535, 2022.

\bibitem[Conley et~al.(2012)Conley, Hansen, and Rossi]{conley2012plausibly}
Timothy~G Conley, Christian~B Hansen, and Peter~E Rossi.
\newblock Plausibly exogenous.
\newblock \emph{Review of Economics and Statistics}, 94\penalty0 (1):\penalty0
  260--272, 2012.

\bibitem[Cui and Tchetgen~Tchetgen(2020)]{cui2020semiparametric}
Yifan Cui and Eric Tchetgen~Tchetgen.
\newblock A semiparametric instrumental variable approach to optimal treatment
  regimes under endogeneity.
\newblock \emph{Journal of the American Statistical Association}, 116\penalty0
  (533):\penalty0 162--173, 2020.

\bibitem[Cui and Tchetgen~Tchetgen(2021)]{cui2019biasaware}
Yifan Cui and Eric Tchetgen~Tchetgen.
\newblock Selective machine learning of doubly robust functionals.
\newblock \emph{Technical Report}, 2021.

\bibitem[Davidian et~al.(2005)Davidian, Tsiatis, and
  Leon]{davidian2005semiparametric}
Marie Davidian, Anastasios~A Tsiatis, and Selene Leon.
\newblock Semiparametric estimation of treatment effect in a pretest--posttest
  study with missing data.
\newblock \emph{Statistical Science}, 20\penalty0 (3):\penalty0 261, 2005.

\bibitem[Dawid(2003)]{dawid2003causal}
A~Philip Dawid.
\newblock {Causal inference using influence diagrams: The problem of partial
  compliance}.
\newblock \emph{Oxford Statistical Science Series}, pages 45--65, 2003.

\bibitem[Didelez et~al.(2010)Didelez, Meng, Sheehan,
  et~al.]{didelez2010assumptions}
Vanessa Didelez, Sha Meng, Nuala~A Sheehan, et~al.
\newblock Assumptions of {IV} methods for observational epidemiology.
\newblock \emph{Statistical Science}, 25\penalty0 (1):\penalty0 22--40, 2010.

\bibitem[Duan and Yin(2017)]{duan2017ensemble}
Xiaogang Duan and Guosheng Yin.
\newblock Ensemble approaches to estimating the population mean with missing
  response.
\newblock \emph{Scandinavian Journal of Statistics}, 44\penalty0 (4):\penalty0
  899--917, 2017.

\bibitem[Engen et~al.(1996)Engen, Gale, and Scholz]{engen1996illusory}
Eric~M Engen, William~G Gale, and John~Karl Scholz.
\newblock The illusory effects of saving incentives on saving.
\newblock \emph{Journal of Economic Perspectives}, 10\penalty0 (4):\penalty0
  113--138, 1996.

\bibitem[Friedman et~al.(2010{\natexlab{a}})Friedman, Hastie, and
  Tibshirani]{friedman2010regularization}
Jerome Friedman, Trevor Hastie, and Rob Tibshirani.
\newblock Regularization paths for generalized linear models via coordinate
  descent.
\newblock \emph{Journal of Statistical Software}, 33\penalty0 (1):\penalty0 1,
  2010{\natexlab{a}}.

\bibitem[Friedman et~al.(2010{\natexlab{b}})Friedman, Hastie, and
  Tibshirani]{glmnet}
Jerome Friedman, Trevor Hastie, and Robert Tibshirani.
\newblock Regularization paths for generalized linear models via coordinate
  descent.
\newblock \emph{Journal of Statistical Software}, 33\penalty0 (1):\penalty0
  1--22, 2010{\natexlab{b}}.

\bibitem[Friedman(2001)]{friedman2001greedy}
Jerome~H Friedman.
\newblock Greedy function approximation: a gradient boosting machine.
\newblock \emph{Annals of Statistics}, pages 1189--1232, 2001.

\bibitem[Goldberger(1972)]{goldberger1972structural}
Arthur~S Goldberger.
\newblock Structural equation methods in the social sciences.
\newblock \emph{Econometrica}, 40\penalty0 (6):\penalty0 979--1001, 1972.

\bibitem[Greenland(2000)]{greenland2000introduction}
Sander Greenland.
\newblock An introduction to instrumental variables for epidemiologists.
\newblock \emph{International Journal of Epidemiology}, 29\penalty0
  (4):\penalty0 722--729, 2000.

\bibitem[Greenwell et~al.(2019)Greenwell, Boehmke, Cunningham, and
  Developers]{gbm}
Brandon Greenwell, Bradley Boehmke, Jay Cunningham, and GBM Developers.
\newblock \emph{gbm: Generalized Boosted Regression Models}, 2019.
\newblock R package version 2.1.5.

\bibitem[Guo et~al.(2018)Guo, Kang, Tony~Cai, and Small]{guo2018confidence}
Zijian Guo, Hyunseung Kang, T~Tony~Cai, and Dylan~S Small.
\newblock Confidence intervals for causal effects with invalid instruments by
  using two-stage hard thresholding with voting.
\newblock \emph{Journal of the Royal Statistical Society: Series B (Statistical
  Methodology)}, 2018.

\bibitem[Han(2008)]{han2008detecting}
Chirok Han.
\newblock Detecting invalid instruments using $l_1$-gmm.
\newblock \emph{Economics Letters}, 101\penalty0 (3):\penalty0 285--287, 2008.

\bibitem[Han(2014)]{han2014multiply}
Peisong Han.
\newblock Multiply robust estimation in regression analysis with missing data.
\newblock \emph{Journal of the American Statistical Association}, 109\penalty0
  (507):\penalty0 1159--1173, 2014.

\bibitem[Han and Wang(2013)]{han2013estimation}
Peisong Han and Lu~Wang.
\newblock Estimation with missing data: beyond double robustness.
\newblock \emph{Biometrika}, 100\penalty0 (2):\penalty0 417--430, 2013.

\bibitem[Hasselman and Hasselman(2018)]{hasselman2018package}
Berend Hasselman and Maintainer~Berend Hasselman.
\newblock Package `nleqslv'.
\newblock 2018.
\newblock R package version 3.2.

\bibitem[Heckman(1997)]{heckman1997instrumental}
James Heckman.
\newblock Instrumental variables: A study of implicit behavioral assumptions
  used in making program evaluations.
\newblock \emph{Journal of Human Resources}, pages 441--462, 1997.

\bibitem[Heckman et~al.(2006)Heckman, Urzua, and
  Vytlacil]{heckman2006understanding}
James~J Heckman, Sergio Urzua, and Edward Vytlacil.
\newblock Understanding instrumental variables in models with essential
  heterogeneity.
\newblock \emph{The Review of Economics and Statistics}, 88\penalty0
  (3):\penalty0 389--432, 2006.

\bibitem[Hern{\'a}n and Robins(2006)]{hernan2006instruments}
Miguel~A Hern{\'a}n and James~M Robins.
\newblock {Instruments for causal inference: An epidemiologist's dream?}
\newblock \emph{Epidemiology}, pages 360--372, 2006.

\bibitem[Imbens(2010)]{imbens2010better}
Guido~W Imbens.
\newblock {Better LATE than nothing: Some comments on Deaton (2009) and Heckman
  and Urzua (2009)}.
\newblock \emph{Journal of Economic Literature}, 48\penalty0 (2):\penalty0
  399--423, 2010.

\bibitem[Imbens(2014)]{imbens2014}
Guido~W. Imbens.
\newblock Instrumental variables: An econometrician's perspective.
\newblock \emph{Statistical Science}, 29\penalty0 (3):\penalty0 323--358, 2014.

\bibitem[Imbens and Angrist(1994)]{angrist1995identification}
Guido~W. Imbens and Joshua~D. Angrist.
\newblock Identification and estimation of local average treatment effects.
\newblock \emph{Econometrica}, 62\penalty0 (2):\penalty0 467--475, 1994.

\bibitem[Imbens and Wooldridge(2009)]{imbens2009recent}
Guido~W Imbens and Jeffrey~M Wooldridge.
\newblock Recent developments in the econometrics of program evaluation.
\newblock \emph{Journal of Economic Literature}, 47\penalty0 (1):\penalty0
  5--86, 2009.

\bibitem[Kang et~al.(2016)Kang, Zhang, Cai, and Small]{kang2016instrumental}
Hyunseung Kang, Anru Zhang, T~Tony Cai, and Dylan~S Small.
\newblock Instrumental variables estimation with some invalid instruments and
  its application to mendelian randomization.
\newblock \emph{Journal of the American Statistical Association}, 111\penalty0
  (513):\penalty0 132--144, 2016.

\bibitem[Klein and Vella(2010)]{klein2010estimating}
Roger Klein and Francis Vella.
\newblock Estimating a class of triangular simultaneous equations models
  without exclusion restrictions.
\newblock \emph{Journal of Econometrics}, 154\penalty0 (2):\penalty0 154--164,
  2010.

\bibitem[Koles{\'a}r et~al.(2015)Koles{\'a}r, Chetty, Friedman, Glaeser, and
  Imbens]{kolesar2015identification}
Michal Koles{\'a}r, Raj Chetty, John Friedman, Edward Glaeser, and Guido~W
  Imbens.
\newblock Identification and inference with many invalid instruments.
\newblock \emph{Journal of Business \& Economic Statistics}, 33\penalty0
  (4):\penalty0 474--484, 2015.

\bibitem[Leon et~al.(2003)Leon, Tsiatis, and Davidian]{leon2003semiparametric}
Selene Leon, Anastasios~A Tsiatis, and Marie Davidian.
\newblock Semiparametric estimation of treatment effect in a pretest-posttest
  study.
\newblock \emph{Biometrics}, 59\penalty0 (4):\penalty0 1046--1055, 2003.

\bibitem[Lewbel(2012)]{lewbel2012using}
Arthur Lewbel.
\newblock Using heteroscedasticity to identify and estimate mismeasured and
  endogenous regressor models.
\newblock \emph{Journal of Business \& Economic Statistics}, 30\penalty0
  (1):\penalty0 67--80, 2012.

\bibitem[Li et~al.(2020)Li, Gu, and Liu]{li2020demystifying}
Wei Li, Yuwen Gu, and Lan Liu.
\newblock Demystifying a class of multiply robust estimators.
\newblock \emph{Biometrika}, 107\penalty0 (4):\penalty0 919--933, 2020.

\bibitem[Liaw et~al.(2002)Liaw, Wiener, et~al.]{liaw2002classification}
Andy Liaw, Matthew Wiener, et~al.
\newblock Classification and regression by randomforest.
\newblock \emph{R news}, 2\penalty0 (3):\penalty0 18--22, 2002.

\bibitem[Malley et~al.(2012)Malley, Kruppa, Dasgupta, Malley, and
  Ziegler]{malley2012probability}
James~D Malley, Jochen Kruppa, Abhijit Dasgupta, Karen~G Malley, and Andreas
  Ziegler.
\newblock Probability machines.
\newblock \emph{Methods of Information in Medicine}, 51\penalty0 (01):\penalty0
  74--81, 2012.

\bibitem[Newey(1990)]{newey1990semiparametric}
Whitney~K Newey.
\newblock Semiparametric efficiency bounds.
\newblock \emph{Journal of Applied Econometrics}, 5\penalty0 (2):\penalty0
  99--135, 1990.

\bibitem[Newey and McFadden(1994)]{newey1994large}
Whitney~K Newey and Daniel McFadden.
\newblock Large sample estimation and hypothesis testing.
\newblock \emph{Handbook of Econometrics}, 4:\penalty0 2111--2245, 1994.

\bibitem[Newey and Windmeijer(2009)]{Newey:2009aa}
Whitney~K. Newey and Frank Windmeijer.
\newblock Generalized method of moments with many weak moment conditions.
\newblock \emph{Econometrica}, 77\penalty0 (3):\penalty0 687--719, 2009.

\bibitem[Neyman(1923)]{neyman1923applications}
Jersey Neyman.
\newblock Sur les applications de la th{\'e}orie des probabilit{\'e}s aux
  experiences agricoles: Essai des principes.
\newblock \emph{Roczniki Nauk Rolniczych}, 10:\penalty0 1--51, 1923.

\bibitem[Neyman(1959)]{neyman1}
Jerzy Neyman.
\newblock Optimal asymptotic tests of composite statistical hypotheses.
\newblock In \emph{Probability and Statistics}, pages 416--44. Wiley, 1959.

\bibitem[Neyman(1979)]{neyman2}
Jerzy Neyman.
\newblock $c(\alpha)$ tests and their use.
\newblock \emph{Sankhya}, pages 1--21, 1979.

\bibitem[Pearl(2009)]{pearl2009causality}
Judea Pearl.
\newblock \emph{Causality}.
\newblock Cambridge University Press, 2009.

\bibitem[Polley et~al.(2021)Polley, LeDell, Kennedy, Lendle, and van~der
  Laan]{polley2021package}
Eric Polley, Erin LeDell, Chris Kennedy, Sam Lendle, and Mark van~der Laan.
\newblock {Package ‘SuperLearner’}, 2021.

\bibitem[Poterba et~al.(1995)Poterba, Venti, and Wise]{poterba1995401}
James~M Poterba, Steven~F Venti, and David~A Wise.
\newblock Do 401 (k) contributions crowd out other personal saving?
\newblock \emph{Journal of Public Economics}, 58\penalty0 (1):\penalty0 1--32,
  1995.

\bibitem[Poterba et~al.(1996)Poterba, Venti, and Wise]{poterba1996retirement}
James~M Poterba, Steven~F Venti, and David~A Wise.
\newblock How retirement saving programs increase saving.
\newblock \emph{Journal of Economic Perspectives}, 10\penalty0 (4):\penalty0
  91--112, 1996.

\bibitem[Richardson and Robins(2013)]{richardson2013single}
Thomas~S Richardson and James~M Robins.
\newblock Single world intervention graphs (swigs): A unification of the
  counterfactual and graphical approaches to causality.
\newblock \emph{Center for the Statistics and the Social Sciences, University
  of Washington Series. Working Paper}, 128\penalty0 (30):\penalty0 2013, 2013.

\bibitem[Richardson et~al.(2017)Richardson, Robins, and
  Wang]{richardson2017modeling}
Thomas~S Richardson, James~M Robins, and Linbo Wang.
\newblock On modeling and estimation for the relative risk and risk difference.
\newblock \emph{Journal of the American Statistical Association}, 112\penalty0
  (519):\penalty0 1121--1130, 2017.

\bibitem[Rigobon(2003)]{rigobon2003identification}
Roberto Rigobon.
\newblock Identification through heteroskedasticity.
\newblock \emph{Review of Economics and Statistics}, 85\penalty0 (4):\penalty0
  777--792, 2003.

\bibitem[Robins et~al.(2009)Robins, Li, Tchetgen, and van~der
  Vaart]{robins2009quadratic}
James Robins, Lingling Li, Eric Tchetgen, and Aad~W van~der Vaart.
\newblock Quadratic semiparametric von mises calculus.
\newblock \emph{Metrika}, 69\penalty0 (2):\penalty0 227--247, 2009.

\bibitem[Robins(1989)]{robins1989analysis}
James~M Robins.
\newblock The analysis of randomized and non-randomized aids treatment trials
  using a new approach to causal inference in longitudinal studies.
\newblock \emph{Health Service Research Methodology: a Focus on AIDS}, pages
  113--159, 1989.

\bibitem[Robins(1994)]{robins1994correcting}
James~M Robins.
\newblock Correcting for non-compliance in randomized trials using structural
  nested mean models.
\newblock \emph{Communications in Statistics-Theory and Methods}, 23\penalty0
  (8):\penalty0 2379--2412, 1994.

\bibitem[Robins and Greenland(1996)]{10.2307/2291630}
James~M. Robins and Sander Greenland.
\newblock Identification of causal effects using instrumental variables:
  Comment.
\newblock \emph{Journal of the American Statistical Association}, 91\penalty0
  (434):\penalty0 456--458, 1996.

\bibitem[Robins and Ritov(1997)]{robins1997toward}
James~M Robins and Ya'acov Ritov.
\newblock Toward a curse of dimensionality appropriate {(CODA)} asymptotic
  theory for semi-parametric models.
\newblock \emph{Statistics in Medicine}, 16\penalty0 (3):\penalty0 285--319,
  1997.

\bibitem[Robins and Rotnitzky(2001)]{robi}
James~M. Robins and Andrea Rotnitzky.
\newblock Comment on ``inference for semiparametric models: Some questions and
  an answer''.
\newblock \emph{Statistica Sinica}, 11:\penalty0 920--936, 2001.

\bibitem[Rubin and van~der Laan(2011)]{rubin2011targeted}
Daniel~B Rubin and Mark~J van~der Laan.
\newblock Targeted {ANCOVA} estimator in {RCTs}.
\newblock In \emph{Targeted Learning}, pages 201--215. Springer, 2011.

\bibitem[Rubin(1974)]{rubin1974estimating}
Donald~B Rubin.
\newblock Estimating causal effects of treatments in randomized and
  nonrandomized studies.
\newblock \emph{Journal of Educational Psychology}, 66\penalty0 (5):\penalty0
  688, 1974.

\bibitem[Rubin(1980)]{rubin1980randomization}
Donald~B Rubin.
\newblock Randomization analysis of experimental data: The {Fisher}
  randomization test comment.
\newblock \emph{Journal of the American Statistical Association}, 75\penalty0
  (371):\penalty0 591--593, 1980.

\bibitem[Rubin(2007)]{rubin2007design}
Donald~B Rubin.
\newblock The design versus the analysis of observational studies for causal
  effects: parallels with the design of randomized trials.
\newblock \emph{Statistics in Medicine}, 26\penalty0 (1):\penalty0 20--36,
  2007.

\bibitem[Shardell and Ferrucci(2016)]{shardell2016instrumental}
Michelle Shardell and Luigi Ferrucci.
\newblock Instrumental variable analysis of multiplicative models with
  potentially invalid instruments.
\newblock \emph{Statistics in Medicine}, 35\penalty0 (29):\penalty0 5430--5447,
  2016.

\bibitem[Shi et~al.(2020)Shi, Miao, Nelson, and
  Tchetgen~Tchetgen]{shi2020multiply}
Xu~Shi, Wang Miao, Jennifer~C Nelson, and Eric~J Tchetgen~Tchetgen.
\newblock Multiply robust causal inference with double-negative control
  adjustment for categorical unmeasured confounding.
\newblock \emph{Journal of the Royal Statistical Society: Series B (Statistical
  Methodology)}, 2020.

\bibitem[Small(2007)]{small2007sensitivity}
Dylan~S Small.
\newblock Sensitivity analysis for instrumental variables regression with
  overidentifying restrictions.
\newblock \emph{Journal of the American Statistical Association}, 102\penalty0
  (479):\penalty0 1049--1058, 2007.

\bibitem[Sun and Tchetgen~Tchetgen(2018)]{sun2018inverse}
BaoLuo Sun and Eric~J Tchetgen~Tchetgen.
\newblock On inverse probability weighting for nonmonotone missing at random
  data.
\newblock \emph{Journal of the American Statistical Association}, 113\penalty0
  (521):\penalty0 369--379, 2018.

\bibitem[Sun et~al.(2018)Sun, Liu, Miao, Wirth, Robins, and
  Tchetgen~Tchetgen]{sun2016semiparametric}
BaoLuo Sun, Lan Liu, Wang Miao, Kathleen Wirth, James Robins, and Eric
  Tchetgen~Tchetgen.
\newblock Semiparametric estimation with data missing not at random using an
  instrumental variable.
\newblock \emph{Statistica Sinica}, 28:\penalty0 1965--1983, 2018.

\bibitem[Swanson et~al.(2018)Swanson, Hern{\'a}n, Miller, Robins, and
  Richardson]{swanson2018partial}
Sonja~A Swanson, Miguel~A Hern{\'a}n, Matthew Miller, James~M Robins, and
  Thomas~S Richardson.
\newblock Partial identification of the average treatment effect using
  instrumental variables: review of methods for binary instruments, treatments,
  and outcomes.
\newblock \emph{Journal of the American Statistical Association}, 113\penalty0
  (522):\penalty0 933--947, 2018.

\bibitem[Tan(2006)]{tan2006distributional}
Zhiqiang Tan.
\newblock A distributional approach for causal inference using propensity
  scores.
\newblock \emph{Journal of the American Statistical Association}, 101\penalty0
  (476):\penalty0 1619--1637, 2006.

\bibitem[Tan(2010)]{tan2010nonparametric}
Zhiqiang Tan.
\newblock Nonparametric likelihood and doubly robust estimating equations for
  marginal and nested structural models.
\newblock \emph{Canadian Journal of Statistics}, 38\penalty0 (4):\penalty0
  609--632, 2010.

\bibitem[Tchetgen~Tchetgen(2009)]{tchetgen2009commentary}
Eric~J Tchetgen~Tchetgen.
\newblock A commentary on g. molenberghs's review of missing data methods.
\newblock \emph{Drug Information Journal}, 43\penalty0 (4):\penalty0 433--435,
  2009.

\bibitem[Tchetgen~Tchetgen and Shpitser(2012)]{tchetgen2012semiparametric}
Eric~J Tchetgen~Tchetgen and Ilya Shpitser.
\newblock Semiparametric theory for causal mediation analysis: efficiency
  bounds, multiple robustness, and sensitivity analysis.
\newblock \emph{Annals of Statistics}, 40\penalty0 (3):\penalty0 1816, 2012.

\bibitem[Tchetgen~Tchetgen et~al.(2009)Tchetgen~Tchetgen, Robins, and
  Rotnitzky]{tchetgen2009doubly}
Eric~J Tchetgen~Tchetgen, James~M Robins, and Andrea Rotnitzky.
\newblock On doubly robust estimation in a semiparametric odds ratio model.
\newblock \emph{Biometrika}, 97\penalty0 (1):\penalty0 171--180, 2009.

\bibitem[Tchetgen~Tchetgen et~al.(2021)Tchetgen~Tchetgen, Sun, and
  Walter]{2019arXiv170907779T}
Eric~J Tchetgen~Tchetgen, BaoLuo Sun, and Stefan Walter.
\newblock The {GENIUS} approach to robust mendelian randomization inference.
\newblock \emph{Statistical Science}, 36\penalty0 (3):\penalty0 443--464, 2021.

\bibitem[Ten~Have et~al.(2008)Ten~Have, Normand, Marcus, Brown, Lavori, and
  Duan]{ten2008intent}
Thomas~R Ten~Have, Sharon Lise~T Normand, Sue~M Marcus, C~Hendricks Brown,
  Philip Lavori, and Naihua Duan.
\newblock Intent-to-treat vs. non-intent-to-treat analyses under treatment
  non-adherence in mental health randomized trials.
\newblock \emph{Psychiatric Annals}, 38\penalty0 (12), 2008.

\bibitem[Tibshirani(1996)]{tibshirani1996regression}
Robert Tibshirani.
\newblock Regression shrinkage and selection via the lasso.
\newblock \emph{Journal of the Royal Statistical Society: Series B
  (Methodological)}, 58\penalty0 (1):\penalty0 267--288, 1996.

\bibitem[Van~der Laan and Rose(2011)]{van2011targeted}
Mark~J Van~der Laan and Sherri Rose.
\newblock \emph{{Targeted Learning: Causal Inference for Observational and
  Experimental Data}}.
\newblock Springer Science \& Business Media, 2011.

\bibitem[Van~der Laan et~al.(2007)Van~der Laan, Polley, and
  Hubbard]{van2007super}
Mark~J Van~der Laan, Eric~C Polley, and Alan~E Hubbard.
\newblock Super {learner}.
\newblock \emph{Statistical Applications in Genetics and Molecular Biology},
  6\penalty0 (1), 2007.

\bibitem[Van~der Vaart et~al.(2006)Van~der Vaart, Dudoit, and van~der
  Laan]{van2006oracle}
Aad~W Van~der Vaart, Sandrine Dudoit, and Mark~J van~der Laan.
\newblock Oracle inequalities for multi-fold cross validation.
\newblock \emph{Statistics and Decisions}, 24\penalty0 (3):\penalty0 351--371,
  2006.

\bibitem[Vansteelandt et~al.(2007)Vansteelandt, Rotnitzky, and
  Robins]{vansteelandt2007estimation}
Stijn Vansteelandt, Andrea Rotnitzky, and James Robins.
\newblock Estimation of regression models for the mean of repeated outcomes
  under nonignorable nonmonotone nonresponse.
\newblock \emph{Biometrika}, 94\penalty0 (4):\penalty0 841--860, 2007.

\bibitem[Vansteelandt et~al.(2008)Vansteelandt, VanderWeele, Tchetgen~Tchetgen,
  and Robins]{vansteelandt2008multiply}
Stijn Vansteelandt, Tyler~J VanderWeele, Eric~J Tchetgen~Tchetgen, and James~M
  Robins.
\newblock Multiply robust inference for statistical interactions.
\newblock \emph{Journal of the American Statistical Association}, 103\penalty0
  (484):\penalty0 1693--1704, 2008.

\bibitem[Vermeulen and Vansteelandt(2015)]{vermeulen2015bias}
Karel Vermeulen and Stijn Vansteelandt.
\newblock Bias-reduced doubly robust estimation.
\newblock \emph{Journal of the American Statistical Association}, 110\penalty0
  (511):\penalty0 1024--1036, 2015.

\bibitem[Wager and Athey(2018)]{wager2018estimation}
Stefan Wager and Susan Athey.
\newblock Estimation and inference of heterogeneous treatment effects using
  random forests.
\newblock \emph{Journal of the American Statistical Association}, 113\penalty0
  (523):\penalty0 1228--1242, 2018.

\bibitem[Wang and Tchetgen~Tchetgen(2018)]{wang2018bounded}
Linbo Wang and Eric Tchetgen~Tchetgen.
\newblock Bounded, efficient and multiply robust estimation of average
  treatment effects using instrumental variables.
\newblock \emph{Journal of the Royal Statistical Society: Series B (Statistical
  Methodology)}, 80\penalty0 (3):\penalty0 531--550, 2018.

\bibitem[Wang et~al.(2018)Wang, Jiang, Zhang, and Small]{wang2018sensitivity}
Xuran Wang, Yang Jiang, Nancy~R Zhang, and Dylan~S Small.
\newblock Sensitivity analysis and power for instrumental variable studies.
\newblock \emph{Biometrics}, 74\penalty0 (4):\penalty0 1150--1160, 2018.

\bibitem[White(1982)]{white1982maximum}
Halbert White.
\newblock Maximum likelihood estimation of misspecified models.
\newblock \emph{Econometrica}, pages 1--25, 1982.

\bibitem[Windmeijer et~al.(2019)Windmeijer, Farbmacher, Davies, and
  Davey~Smith]{windmeijer2018use}
Frank Windmeijer, Helmut Farbmacher, Neil Davies, and George Davey~Smith.
\newblock On the use of the lasso for instrumental variables estimation with
  some invalid instruments.
\newblock \emph{Journal of the American Statistical Association}, 114\penalty0
  (527):\penalty0 1339--1350, 2019.

\bibitem[Wooldridge(2010)]{wooldridge2010econometric}
Jeffrey~M Wooldridge.
\newblock \emph{Econometric {Analysis} of {Cross} {Section} and {Panel}
  {Data}}.
\newblock MIT Press, 2010.

\bibitem[Wright and Ziegler(2017)]{ranger}
Marvin~N. Wright and Andreas Ziegler.
\newblock {ranger}: A fast implementation of random forests for high
  dimensional data in {C++} and {R}.
\newblock \emph{Journal of Statistical Software}, 77\penalty0 (1):\penalty0
  1--17, 2017.
\newblock \doi{10.18637/jss.v077.i01}.

\bibitem[Wright(1928)]{wright1928tariff}
Philip~G Wright.
\newblock \emph{{Tariff on Animal and Vegetable Oils}}.
\newblock Macmillan Company, New York, 1928.

\bibitem[Ye et~al.(2021)Ye, Liu, Sun, and Tchetgen]{ye2021geniusmawii}
Ting Ye, Zhonghua Liu, Baoluo Sun, and Eric~Tchetgen Tchetgen.
\newblock {GENIUS-MAWII}: For robust mendelian randomization with many weak
  invalid instruments.
\newblock \emph{Technical Report}, 2021.

\bibitem[Zheng and Van Der~Laan(2010)]{zheng2010asymptotic}
Wenjing Zheng and Mark~J Van Der~Laan.
\newblock Asymptotic theory for cross-validated targeted maximum likelihood
  estimation.
\newblock \emph{Technical Report}, 2010.

\end{thebibliography}

\end{document}